\documentclass[a4paper,12pt]{article}

\usepackage{amsmath}
\usepackage{array}
\usepackage{amssymb}
\usepackage{amsthm}
\usepackage{mathrsfs}
\usepackage[T1]{fontenc}
\usepackage{enumerate}
\usepackage{fullpage}
\usepackage{multirow}
\usepackage{color}

\def\eqn{equation}
\def\cond{condition}

\def\tfn{transformation}

\def\sm{sigma model}
\def\pl{Poisson--Lie}

\def\dd{Drinfeld double}
\def\mt{Manin triple}

\def\4diml{four-dimensional}
\def\bkg{background}

\def\-1{^{-1}}

\def\coor{coordinate}
\def\real{\mathbb{R}}

\def\cd{{\mathfrak d}}
\def\cg{{\mathfrak g}}
\def\tcg{\tilde{\mathfrak g}}

\def\wt{\tilde}

\def\sm{sigma model}
\def\PL{Poisson--Lie }

\def\pltd{Poisson--Lie T-dualit}

\newcommand{\unit}{\mathbf{1}}

\newcommand{\D}{\mathscr{D}}
\newcommand{\G}{\mathscr{G}}
\newcommand{\tG}{\widetilde{\mathscr{G}}}

\begin{document}
\title{Classification of standard Manin triples in dimension 4+4}
\author{Ladislav Hlavat\'y\footnote{ladislav.hlavaty@fjfi.cvut.cz}, Petr Novotn\'y\footnote{petr.novotny@fjfi.cvut.cz}
\\ {\em Faculty of Nuclear Sciences and Physical Engineering,}
\\ {\em Czech Technical University in Prague,}
\\ {\em B\v rehov\' a 7, 115 19, Prague 1,}
\\ {\em Czech Republic}
\and
Ivo Petr\footnote{ivo.petr@fit.cvut.cz}
\\ {\em Faculty of Information Technology,}
\\ {\em Czech Technical University in Prague,}
\\ {\em Th\' akurova 9, 160 00, Prague 6,}
\\ {\em Czech Republic}
}
\maketitle

\abstract{Four- and six-dimensional Drinfeld doubles were classified in the past in terms of Manin triples. We provide an important step towards the classification of eight-dimensional Drinfeld doubles by presenting an extensive list of Manin triples formed by pairs of four-dimensional Lie algebras. Due to the high complexity of the classification we focus on Manin triples formed by algebras in a certain standard form. The list contains 188 non-isomorphic Manin triples plus their duals. 

To apply the results, we construct several four-dimensional WZW models on non-semisimple Lie groups. Some of the WZW models are known from the literature, but new cases are presented as well. As a consequence of the construction method, the WZW models are Poisson--Lie dualizable.}

\tableofcontents


\section{Introduction}

Drinfeld doubles and Manin triples are crucial objects in the study of dualities in string theory. The well-known Abelian T-duality \cite{buscher:ssbfe} and non-Abelian T-duality \cite{rocver} rely on the presence of (non-)Abelian symmetries of the \bkg\ fields. However, as the dual model may not always have the required symmetries, it might be impossible to obtain the original model from the dual one. The introduction of \PL T-duality --- a generalization of non-Abelian T-duality --- allowed researchers to treat mutually dual models on an equal footing by requiring \PL symmetries \cite{klise,klim:proc,unge:pltp}. The key to this generalization lies in the structure of the \dd. By providing a unified setting for both the original and dual sigma models, the Drinfeld double enables the construction of dual and plural geometries and the consistent formulation of dual and plural field theories.  

Finding \PL duals of a particular background may be highly non-trivial, but one may choose a different approach. The algebra of the \dd\ can be decomposed in several ways into pairs of Lie algebras forming Manin triples. For each Manin triple it is possible to construct a dualizable or pluralizable sigma model on the corresponding Lie group. A number of physically interesting \bkg s were obtained in this way and studied, including plane-parallel waves \cite{hlatur,hlapetr}, Bianchi cosmologies \cite{dualnikorejci,plidb,hlape:bianchisugra}, or $AdS_3\times S^3$ and $AdS_5\times S^5$ \bkg s \cite{saka2,eghbaliAdS}. Due to the close relation between Manin triples and Lie bialgebras the structure of a Manin triple also plays an important role in the study of integrable models and their deformations \cite{klim:int, stijn1, araujo, dehath, borwulff}.

Classification of four- and six-dimensional \dd s and Manin triples carried out in Refs. \cite{LHLS:4ddd,gom:ctd,hlasno:mtriples,snohla:DD} allowed systematic construction of the aforementioned models. The models are usually built on two- or three- dimensional Lie groups. In order to obtain a physical \bkg\ on a manifold of higher dimension, one has to include spectator fields. There are examples of models on four-dimensional Lie groups \cite{hlatur,eghbali11,saka:wzw} where T-duality was used and eight-dimensional \dd s were exploited. However, there are not many of them and they mostly reduce to non-Abelian T-duality. The classification of eight-dimensional \dd s would allow us to study the full \PL T-duality and plurality of these models. For that we have to classify the corresponding \mt s first.

As we shall see in the following sections, searching for Manin triples is essentially a matter of solving systems of quadratic equations following from Jacobi identities. The higher the dimension, the more variables --- the Lie algebra structure constants --- appear in the equations. For Manin triples composed of four-dimensional algebras it is quite challenging to provide a complete classification. In dimension 3+3 in Refs. \cite{hlasno:mtriples,snohla:DD} it turned out that for most of the Manin triples a representative (up to isomorphisms of Manin triples) with certain ``standard'' form (see below) can be chosen. Therefore, we shall use the classification of four-dimensional Lie algebras given in Refs. \cite{Mubarakzyan,pawiza,snowin:kniha}, and look for pairs of these algebras forming Manin triples in the way explained below. Besides the algebras presented in the Refs. \cite{pawiza,snowin:kniha}, we also consider permutations and scalings of their generators. Although these transformations are just isomorphisms of the four-dimensional algebras, they may lead to different Manin triples. In the large number of Manin triples obtained in this way we identify non-isomorphic ones. As the main result of the paper we present a list of ``standard'' \mt s formed by pairs of four-dimensional Lie algebras. The list contains $188$ non-isomorphic Manin triples that in several cases depend on real parameters. In each case a dual \mt\ can be obtained by switching the particular algebras. After we have published our results on ArXiv, another approach investigating solutions of Jacobi identities for indecomposable four-dimensional algebras appeared in Ref. \cite{Musaev}.

The presented list of Manin triples can be used e.g. for systematic construction of dualizable sigma model \bkg s and finding new solutions to (generalized) supergravity equations \cite{sibylla, bdh}. Particularly interesting models discussed frequently in the literature are the WZW models, which give exact string backgrounds. Four-dimensional WZW models on non-semisimple groups corresponding to centrally extended Euclidean algebra $E^c_2 \cong s_{4,7}$ and Heisenberg algebra $H_4 \cong s_{4,6}$  were constructed in Refs. \cite{NW, kehame,eghbali,rezaseph:H4}. Having Manin triples containing these algebras, we construct WZW models as dualizable sigma models. We also show how the \dd\ $(s_{4,6}|s_{2,1}\oplus A_2; \operatorname{P1})$ containing algebra $H_4$ decomposes into different \mt s thus allowing us to study not only dual but also plural models to the WZW model on $H_4$. A previous  study of the WZW models in the context of \pltd y was carried out in Refs. \cite{saka:wzw, qpltwzw, sasa}.

We begin the discussion by reviewing the notion of \dd\ and \mt\ in Section \ref{sec:dd_mt}. The method of classification of \mt s is described in Section \ref{sec:method} and the results are summarized in Sections \ref{sec:no_param}, \ref{sec:one_param}, and \ref{sec:two_param}. In Section \ref{sec:dd_class} we show how Manin triples forming the same \dd\ can be identified, and in Section \ref{sec:wzw} we construct \PL dualizable WZW models.

\section{\dd s and Manin triples}\label{sec:dd_mt}

\dd \ $\D=(\G|\tG)$ is a real $2D$-dimensional Lie group whose Lie algebra $\cd$  decomposes as a vector space into the direct sum of two $D$-dimensional subalgebras $\cg$ and $\tcg$. Moreover, the Lie algebra $\cd$ is equipped with an ad-invariant non-degenerate symmetric bilinear form $\langle . , . \rangle$ and the subalgebras $\cg$ and $\tcg$ are maximally isotropic with respect to $\langle . , . \rangle$.  These three algebras form a Manin triple $(\cd, \cg, \tcg)$.

One can choose mutually dual bases $T_i \in \cg,\ \widetilde{T}^i \in \tcg$, $i=1, \ldots, D,$ such that the generators satisfy relations
\begin{align}
\label{dual_bases}
\langle T_i, T_j \rangle & = 0, & \langle \widetilde{T}^i, \widetilde{T}^j \rangle &= 0, & \langle T_i, \widetilde{T}^j \rangle & = \delta_i^j.
\end{align}
Due to the ad-invariance of the bilinear form $\langle . , . \rangle$ the algebraic structure of the Manin triple is given by commutation relations of the subalgebras $\cg$ and $\tcg$
\begin{equation}
\label{commutation_on_d}
[T_i,T_j]=f_{ij}{}^k T_k, \qquad [\widetilde T^i, \widetilde T^j]=\wt f^{ij}{}_k \widetilde T^k, \qquad [T_i,\widetilde T^j]=f_{ki}{}^j \widetilde T^k + \wt f^{jk}{}_i T_k.
\end{equation}
Since $\cd, \cg, \tcg$ are Lie algebras, the structure constants are antisymmetric, i.e. 
\begin{equation}\label{sc_antisym}
{f}_{kl}{}^{m}=-{f}_{lk}{}^{m}, \qquad \wt{f}^{kl}{}_{m} = -\wt{f}^{lk}{}_{m},
\end{equation}
and satisfy conditions following from Jacobi identities
\begin{equation}\label{g_sc}
{f}_{kl}{}^{m} \, {f}_{ij}{}^{l} + {f}_{il}{}^{m} \, {f}_{jk}{}^{l} + {f}_{jl}{}^{m} \, {f}_{ki}{}^{l}=0,
\end{equation}
\begin{equation}\label{gtilde_sc}
\wt{f}^{kl}{}_{m} \, \wt{f}^{ij}{}_{l} + \wt{f}^{il}{}_{m} \, \wt{f}^{jk}{}_{l} + \wt{f}^{jl}{}_{m} \, \wt{f}^{ki}{}_{l}=0,
\end{equation}
\begin{equation}\label{dd_sc}
\wt{f}^{jk}{}_{l} \, f_{mi}{}^{l}+\wt{f}^{kl}{}_{m} \, f_{li}{}^{j}+\wt{f}^{jl}{}_{i} \, f_{lm}{}^{k}+\wt{f}^{jl}{}_{m} \, f_{il}{}^{k}+\wt{f}^{lk}{}_{i} \, f_{lm}{}^{j}=0.
\end{equation}
Any triple of Lie algebras whose structure constants satisfy equations \eqref{sc_antisym}--\eqref{dd_sc} forms a Manin triple.

Let us consider an automorphism $A$ of the vector space given by $\cg$. The choice \eqref{dual_bases} is invariant with respect to the transformation 
\begin{equation}\label{iso_A}
T'_i = A_i{}^{k} T_k, \qquad \widetilde{T}'^i = \widetilde T^k(A^{-1})_k{}^{i}.
\end{equation}
Under this transformation, the structure constants change as
\begin{equation}
f'_{ij}{}^k = A_i{}^{l} A_j{}^{m} f_{lm}{}^n (A^{-1})_n{}^k, \qquad \wt f'^{ij}{}_k = (A\-1)_{l}{}^i (A\-1)_{m}{}^j \wt f^{lm}{}_n A_k{}^n. \label{iso}
\end{equation}
\mt s $MT = (\cd, \cg, \tcg)$ and $MT' = (\cd', \cg', \tcg')$ are considered \emph{isomorphic} if and only if there is a $4\times 4$ matrix $ A_i{}^{j}$ such that their structure constants are related by \eqref{iso}.

When searching for Manin triples $(\cd, \cg, \tcg)$ we do not have to solve the full system of equations \eqref{sc_antisym}--\eqref{dd_sc}. Low-dimensional real Lie algebras were classified in Refs. \cite{Mubarakzyan,pawiza,snowin:kniha}, so we can choose the algebra $\cg$ and ask what are the possible dual algebras $\tcg$. Equations \eqref{g_sc} are then satisfied and \eqref{dd_sc} become linear. However, we still face the problem of solving the non-linear system of equations \eqref{gtilde_sc}. This approach was chosen in Ref. \cite{hlasno:mtriples}, where \mt s in dimension 3+3 were classified. In dimension 4+4 solving \eqref{gtilde_sc} is computationally demanding and even if the system can be solved, it is difficult to identify all non-isomorphic \mt s and give a full classification. Therefore, we have chosen a different approach.

\section{Method of classification of \mt s}\label{sec:method}

There are 25 non-isomorphic real Lie algebras $A_{4,j}$, $j = 0, \ldots , 24$ of dimension four. Five of the algebras depend on a real parameter $a$ and two of them depend on two real parameters denoted $a,b$. Twelve of the algebras are decomposable. The notation of the indecomposable algebras follows the book \cite{snowin:kniha}. For the notation of the decomposable algebras we use the well-known Bianchi classification. Their definitions in terms of Lie products of their generators can be found in the Appendix in Tables \ref{2dimalras}--\ref{4dimalras}.

By $A_{4,0}$ we denote the Abelian algebra $A_4$, which forms a \mt\ with any other algebra $A_{4,j}$. For the sake of brevity, these 25 semi-Abelian \mt s are omitted from the list given in Sections \ref{sec:no_param}, \ref{sec:one_param} and \ref{sec:two_param}. For each \mt\ $(\cd, \cg, \tcg)$ one can easily obtain a dual \mt\ by simply switching the roles of $\cg$ and $\tcg$. We do not list these dual \mt s either.
 
In this paper we classify \mt s $(\cd, \cg, \tcg)$ that we call \emph{standard}. These are formed by pairs of four-dimensional Lie algebras, where $\cg$ has the form presented in the Tables \ref{2dimalras}--\ref{4dimalras}, while the structure constants of $\tcg$ (called dual algebra or coalgebra) are modified by permutations and overall scalings of the generators. This gives about 14 000 candidates for the standard \mt s but, as we will see, the final count is less than 200 cases plus their duals. The standard \mt s in the 3+3-dimensional classification in Ref. \cite{hlasno:mtriples} do not cover all the cases but the vast majority. The non-standard \mt s will be investigated in a future paper.

To find the classification of standard \mt s we have performed the following steps:
\begin{enumerate}
\item First, we check the Jacobi identities \eqref{dd_sc} for each pair of the algebras\footnote{The Jacobi identities \eqref{g_sc} and \eqref{gtilde_sc} are satisfied since $A_{4,j}$ are Lie algebras.} $\mathfrak{g},\tilde{\mathfrak{g}}$. Algebras $\mathfrak{g}= A_{4,j}, j=1,\ldots, 24,$ are always considered in the form given in Tables \ref{2dimalras}--\ref{4dimalras}. For algebras $\tcg$ we take scaled and permuted versions of $A_{4,k}$. Here $k = 1, \ldots, j$ to avoid double counting of \mt s. In this step we exclude most of the candidates for $\tilde{\mathfrak{g}}$, and for the algebras with parameters we may get restrictions on their values. In this way we obtain a preliminary list of standard \mt s, where some of them are isomorphic.

For example, for the algebra $\cg = s_{4,7}$, that is the $E_2^c$ used in the Nappi--Witten WZW model \cite{NW}, we get only eight possibly isomorphic \mt s from 600 candidates.

\item In the second step, we investigate isomorphisms among the \mt s from the preliminary list obtained in the first step. For solving \eqn s \eqref{iso} {we use computer algebra systems}. The Lie algebras $\mathfrak{g}$ and $\mathfrak{g'}$ in the \mt s $MT, MT'$ are already classified up to isomorphisms, and we can set $f'_{ij}{}^k = f_{ij}{}^k $. In other words, we look for isomorphisms of Lie algebras $\tilde{\mathfrak{g}}, \tilde{\mathfrak{g}}'$ that are automorphisms\footnote{Let us note that those dual algebras $\tilde{\mathfrak{g}}$ that differ only by permutations of the bases are of course isomorphic but this isomorphism is in general not an automorphism of $\cg$.} of the algebra $ \mathfrak{g}$. Identification of isomorphic \mt s reduces the preliminary list and, in certain cases, further restricts the values of free parameters of the algebras.

For example, of the eleven \mt s containing $\cg = s_{4,7}$ obtained in the first step only three are non-isomorphic. They can be found in Section \ref{s47}.

\item In the first step we allowed scalings of the generators of the algebra $\tcg$. Therefore, in the third step we determine which scaling factors $\beta$ give non-isomorphic \mt s. Often, the isomorphisms of \mt s can be used to reduce $\beta$ to $1$. When this is not possible, the scaling factor appears in the classification as a new parameter in one of the three forms:
$$
\beta \in \mathbb{R} \setminus \{0\}, \quad \text{ or } \quad \gamma := \beta > 0, \quad \text{ or } \quad \epsilon := \beta = \pm 1.
$$
\end{enumerate}

Applying this method of classification we finally obtain $188$ non-isomorphic\footnote{At present we do not have rigorous proofs that the displayed \mt s are non-isomorphic. We have independently used computer algebra systems Maple and Wolfram Mathematica to find the isomorphisms and the results were checked against each other.} $(4+4)$-dimensional standard \mt s. For their identification we have chosen the notation
\begin{equation}\label{notation}
(A_{4,j}; \text{par}_j|A_{4,k}; \text{par}_k; \text{permutation}_k , \text{scaling factor } \beta)
\end{equation}
where $A_{4,j}, A_{4,k}$ are the four-dimensional algebras (both decomposable and indecomposable) described in the Appendix. The ordering, given in the Appendix, was chosen such that algebras $A_{4,j}$ with fewer parameters come first. The parameters $\text{par}_j, \text{par}_k$ of the algebras $\cg, \tcg$ are  either empty or $a, b$. Similarly, the scaling factor is either empty (if $\beta = 1$), or $\beta, \gamma, \epsilon$. The ordering of permutations is standard but we include it in the Appendix for completeness.

For brevity, we omit semi-Abelian \mt s and display only the \mt s where  $k\leq j$. Beside these there are  \mt s 
$$
(A_{4,k}; \text{par}_k | A_{4,j}; \text{par}_j; \text{permutation}_j , \text{scaling factor } 1/\beta)  $$
for $k > j$ dual to \eqref{notation}, where $\text{permutation}_j$ is the inverse of $\text{permutation}_k$. Usually, the dual \mt s are not isomorphic to their counterparts, but some of them are self-dual.

\section{Manin triples for algebras without parameters}\label{sec:no_param}

\subsection{Manin triples with $\mathfrak{g}=s_{2,1}\oplus s_{2,1}$}
Lie products of algebra $\mathfrak{g}=s_{2,1}\oplus s_{2,1}$:
$$[T_1, T_2]= T_2, \quad [T_3, T_4]= T_4.$$
Manin triples, permutations and Lie products of algebra $\tilde{\mathfrak{g}}$:
\begin{enumerate}
\item $(s_{2,1}\oplus s_{2,1}|s_{2,1}\oplus s_{2,1}; \operatorname{P1}, \beta\,)$

$$[\widetilde{T}^1, \widetilde{T}^2]=\beta \widetilde{T}^2, \quad [\widetilde{T}^3, \widetilde{T}^4]=\beta \widetilde{T}^4.$$
\item $(s_{2,1}\oplus s_{2,1}|s_{2,1}\oplus s_{2,1}; \operatorname{P2}, \beta\,)$

$$[\widetilde{T}^1, \widetilde{T}^2]=\beta \widetilde{T}^2, \quad [\widetilde{T}^3, \widetilde{T}^4]=-\beta \widetilde{T}^3.$$
\item $(s_{2,1}\oplus s_{2,1}|s_{2,1}\oplus s_{2,1}; \operatorname{P8})$

$$[\widetilde{T}^1, \widetilde{T}^2]=- \widetilde{T}^1, \quad [\widetilde{T}^3, \widetilde{T}^4]=- \widetilde{T}^3.$$
\end{enumerate}
\subsection{Manin triples with $\mathfrak{g}=B2\oplus A_1$}
Lie products of algebra $\mathfrak{g}=B2\oplus A_1$:
$$[T_2, T_3]= T_1.$$
Manin triples, permutations and Lie products of algebra $\tilde{\mathfrak{g}}$:
\begin{enumerate}
\item $(B2\oplus A_1|B2\oplus A_1; \operatorname{P7}, \epsilon\,)$

$$[\widetilde{T}^1, \widetilde{T}^3]=\epsilon \widetilde{T}^2.$$
\item $(B2\oplus A_1|B2\oplus A_1; \operatorname{P8})$

$$[\widetilde{T}^1, \widetilde{T}^4]= \widetilde{T}^2.$$
\item $(B2\oplus A_1|B2\oplus A_1; \operatorname{P10})$

$$[\widetilde{T}^1, \widetilde{T}^2]= \widetilde{T}^4.$$
\item $(B2\oplus A_1|B2\oplus A_1; \operatorname{P19})$

$$[\widetilde{T}^3, \widetilde{T}^4]= \widetilde{T}^2.$$
\end{enumerate}
\subsection{Manin triples with {$\mathfrak{g}=s_{2,1}\oplus A_2 \cong B3\oplus A_1$}}
Lie products of algebra {$\mathfrak{g}=s_{2,1}\oplus A_2$}:
$$[T_1, T_2]= T_2.$$
Manin triples, permutations and Lie products of algebra $\tilde{\mathfrak{g}}$:
\begin{enumerate}
\item $(s_{2,1}\oplus A_2|s_{2,1}\oplus s_{2,1}; \operatorname{P1}, \beta\,)$

$$[\widetilde{T}^1, \widetilde{T}^2]=\beta \widetilde{T}^2, \quad [\widetilde{T}^3, \widetilde{T}^4]=\beta \widetilde{T}^4.$$
\item $(s_{2,1}\oplus A_2|s_{2,1}\oplus s_{2,1}; \operatorname{P7})$

$$[\widetilde{T}^1, \widetilde{T}^2]=- \widetilde{T}^1, \quad [\widetilde{T}^3, \widetilde{T}^4]= \widetilde{T}^4.$$
\item $(s_{2,1}\oplus A_2|B2\oplus A_1; \operatorname{P1})$

$$[\widetilde{T}^2, \widetilde{T}^3]= \widetilde{T}^1.$$
\item $(s_{2,1}\oplus A_2|B2\oplus A_1; \operatorname{P5})$

$$[\widetilde{T}^3, \widetilde{T}^4]= \widetilde{T}^1.$$
\item $(s_{2,1}\oplus A_2|s_{2,1}\oplus A_2; \operatorname{P1}, \beta\,)$

$$[\widetilde{T}^1, \widetilde{T}^2]=\beta \widetilde{T}^2.$$
\item $(s_{2,1}\oplus A_2|s_{2,1}\oplus A_2; \operatorname{P7})$

$$[\widetilde{T}^1, \widetilde{T}^2]=- \widetilde{T}^1.$$
\item $(s_{2,1}\oplus A_2|s_{2,1}\oplus A_2; \operatorname{P15})$

$$[\widetilde{T}^2, \widetilde{T}^3]=- \widetilde{T}^2.$$
\item $(s_{2,1}\oplus A_2|s_{2,1}\oplus A_2; \operatorname{P17})$

$$[\widetilde{T}^3, \widetilde{T}^4]= \widetilde{T}^4.$$
\end{enumerate}
\subsection{Manin triples with $\mathfrak{g}=B4\oplus A_1$}
Lie products of algebra $\mathfrak{g}=B4\oplus A_1$:
$$[T_1, T_2]=- T_2+ T_3, \quad [T_1, T_3]=- T_3.$$
Manin triples, permutations and Lie products of algebra $\tilde{\mathfrak{g}}$:
\begin{enumerate}
\item $(B4\oplus A_1|B2\oplus A_1; \operatorname{P1}, \epsilon\,)$

$$[\widetilde{T}^2, \widetilde{T}^3]=\epsilon \widetilde{T}^1.$$
\item $(B4\oplus A_1|B2\oplus A_1; \operatorname{P2})$

$$[\widetilde{T}^2, \widetilde{T}^4]= \widetilde{T}^1.$$
\item $(B4\oplus A_1|B2\oplus A_1; \operatorname{P5})$

$$[\widetilde{T}^3, \widetilde{T}^4]= \widetilde{T}^1.$$
\item $(B4\oplus A_1|B2\oplus A_1; \operatorname{P7}, \beta\,)$

$$[\widetilde{T}^1, \widetilde{T}^3]=\beta \widetilde{T}^2.$$
\item $(B4\oplus A_1|B2\oplus A_1; \operatorname{P19})$

$$[\widetilde{T}^3, \widetilde{T}^4]= \widetilde{T}^2.$$
\item $(B4\oplus A_1|B4\oplus A_1; \operatorname{P24})$

$$[\widetilde{T}^2, \widetilde{T}^4]= \widetilde{T}^2, \quad [\widetilde{T}^3, \widetilde{T}^4]=- \widetilde{T}^2+ \widetilde{T}^3.$$
\end{enumerate}
\subsection{Manin triples with $\mathfrak{g}=B5\oplus A_1$}
Lie products of algebra $\mathfrak{g}=B5\oplus A_1$:
$$[T_1, T_2]=- T_2, \quad [T_1, T_3]=- T_3.$$
Manin triples, permutations and Lie products of algebra $\tilde{\mathfrak{g}}$:
\begin{enumerate}
\item $(B5\oplus A_1|B2\oplus A_1; \operatorname{P1})$

$$[\widetilde{T}^2, \widetilde{T}^3]= \widetilde{T}^1.$$
\item $(B5\oplus A_1|B2\oplus A_1; \operatorname{P2})$

$$[\widetilde{T}^2, \widetilde{T}^4]= \widetilde{T}^1.$$
\item $(B5\oplus A_1|B2\oplus A_1; \operatorname{P7})$

$$[\widetilde{T}^1, \widetilde{T}^3]= \widetilde{T}^2.$$
\item $(B5\oplus A_1|B2\oplus A_1; \operatorname{P19})$

$$[\widetilde{T}^3, \widetilde{T}^4]= \widetilde{T}^2.$$
\item $(B5\oplus A_1|s_{2,1}\oplus A_2; \operatorname{P16})$

$$[\widetilde{T}^2, \widetilde{T}^4]=- \widetilde{T}^2.$$
\item $(B5\oplus A_1|B4\oplus A_1; \operatorname{P22})$

$$[\widetilde{T}^2, \widetilde{T}^4]= \widetilde{T}^2- \widetilde{T}^3, \quad [\widetilde{T}^3, \widetilde{T}^4]= \widetilde{T}^3.$$
\item $(B5\oplus A_1|B5\oplus A_1; \operatorname{P22})$

$$[\widetilde{T}^2, \widetilde{T}^4]= \widetilde{T}^2, \quad [\widetilde{T}^3, \widetilde{T}^4]= \widetilde{T}^3.$$
\end{enumerate}
\subsection{Manin triples with $\mathfrak{g}=B6_0\oplus A_1$}
Commutation relations of algebra $\mathfrak{g}=B6_0\oplus A_1$:
$$[T_1, T_3]= T_2, \quad [T_2, T_3]= T_1.$$
Manin triples, permutations and Lie products of algebra $\tilde{\mathfrak{g}}$:
\begin{enumerate}
\item $(B6_0\oplus A_1|B2\oplus A_1; \operatorname{P9})$

$$[\widetilde{T}^1, \widetilde{T}^2]= \widetilde{T}^3.$$
\item $(B6_0\oplus A_1|B2\oplus A_1; \operatorname{P10})$

$$[\widetilde{T}^1, \widetilde{T}^2]= \widetilde{T}^4.$$
\item $(B6_0\oplus A_1|B2\oplus A_1; \operatorname{P11})$

$$[\widetilde{T}^1, \widetilde{T}^4]= \widetilde{T}^3.$$
\item $(B6_0\oplus A_1|B4\oplus A_1; \operatorname{P1}, \beta\,)$

$$[\widetilde{T}^1, \widetilde{T}^2]=-\beta \widetilde{T}^2+\beta \widetilde{T}^3, \quad [\widetilde{T}^1, \widetilde{T}^3]=-\beta \widetilde{T}^3.$$
\item $(B6_0\oplus A_1|B5\oplus A_1; \operatorname{P1})$

$$[\widetilde{T}^1, \widetilde{T}^2]=- \widetilde{T}^2, \quad [\widetilde{T}^1, \widetilde{T}^3]=- \widetilde{T}^3.$$
\item $(B6_0\oplus A_1|B5\oplus A_1; \operatorname{P9}, \gamma\,)$

$$[\widetilde{T}^1, \widetilde{T}^3]=\gamma \widetilde{T}^1, \quad [\widetilde{T}^2, \widetilde{T}^3]=\gamma \widetilde{T}^2.$$
\item $(B6_0\oplus A_1|B5\oplus A_1; \operatorname{P10})$

$$[\widetilde{T}^1, \widetilde{T}^4]= \widetilde{T}^1, \quad [\widetilde{T}^2, \widetilde{T}^4]= \widetilde{T}^2.$$
\item $(B6_0\oplus A_1|B6_0\oplus A_1; \operatorname{P2})$

$$[\widetilde{T}^1, \widetilde{T}^4]= \widetilde{T}^2, \quad [\widetilde{T}^2, \widetilde{T}^4]= \widetilde{T}^1.$$
\end{enumerate}
\subsection{Manin triples with $\mathfrak{g}=B7_0\oplus A_1$}
Lie products of algebra $\mathfrak{g}=B7_0\oplus A_1$:
$$[T_1, T_3]=- T_2, \quad [T_2, T_3]= T_1.$$
Manin triples, permutations and Lie products of algebra $\tilde{\mathfrak{g}}$:
\begin{enumerate}
\item $(B7_0\oplus A_1|B2\oplus A_1; \operatorname{P9}, \epsilon\,)$

$$[\widetilde{T}^1, \widetilde{T}^2]=\epsilon \widetilde{T}^3.$$
\item $(B7_0\oplus A_1|B2\oplus A_1; \operatorname{P10})$

$$[\widetilde{T}^1, \widetilde{T}^2]= \widetilde{T}^4.$$
\item $(B7_0\oplus A_1|B2\oplus A_1; \operatorname{P11})$

$$[\widetilde{T}^1, \widetilde{T}^4]= \widetilde{T}^3.$$
\item $(B7_0\oplus A_1|B4\oplus A_1; \operatorname{P1}, \beta\,)$

$$[\widetilde{T}^1, \widetilde{T}^2]=-\beta \widetilde{T}^2+\beta \widetilde{T}^3, \quad [\widetilde{T}^1, \widetilde{T}^3]=-\beta \widetilde{T}^3.$$
\item $(B7_0\oplus A_1|B5\oplus A_1; \operatorname{P1})$

$$[\widetilde{T}^1, \widetilde{T}^2]=- \widetilde{T}^2, \quad [\widetilde{T}^1, \widetilde{T}^3]=- \widetilde{T}^3.$$
\item $(B7_0\oplus A_1|B5\oplus A_1; \operatorname{P9}, \gamma\,)$

$$[\widetilde{T}^1, \widetilde{T}^3]=\gamma \widetilde{T}^1, \quad [\widetilde{T}^2, \widetilde{T}^3]=\gamma \widetilde{T}^2.$$
\item $(B7_0\oplus A_1|B5\oplus A_1; \operatorname{P10})$

$$[\widetilde{T}^1, \widetilde{T}^4]= \widetilde{T}^1, \quad [\widetilde{T}^2, \widetilde{T}^4]= \widetilde{T}^2.$$
\item $(B7_0\oplus A_1|B7_0\oplus A_1; \operatorname{P2})$

$$[\widetilde{T}^1, \widetilde{T}^4]=- \widetilde{T}^2, \quad [\widetilde{T}^2, \widetilde{T}^4]= \widetilde{T}^1.$$
\end{enumerate}
\subsection{Manin triples with $\mathfrak{g}=B8\oplus A_1$}
Lie products of algebra $\mathfrak{g}=B8\oplus A_1$:
$$[T_1, T_2]=- T_3, \quad [T_1, T_3]=- T_2, \quad [T_2, T_3]= T_1.$$
Manin triples, permutations and Lie products of algebra $\tilde{\mathfrak{g}}$:
\begin{enumerate}
\item $(B8\oplus A_1|B5\oplus A_1; \operatorname{P1}, \gamma\,)$

$$[\widetilde{T}^1, \widetilde{T}^2]=-\gamma \widetilde{T}^2, \quad [\widetilde{T}^1, \widetilde{T}^3]=-\gamma \widetilde{T}^3.$$
\item $(B8\oplus A_1|B5\oplus A_1; \operatorname{P9}, \gamma\,)$

$$[\widetilde{T}^1, \widetilde{T}^3]=\gamma \widetilde{T}^1, \quad [\widetilde{T}^2, \widetilde{T}^3]=\gamma \widetilde{T}^2.$$
\item $(B8\oplus A_1|B6_0\oplus A_1; \operatorname{P5})$

$$[\widetilde{T}^1, \widetilde{T}^4]= \widetilde{T}^3, \quad [\widetilde{T}^3, \widetilde{T}^4]= \widetilde{T}^1.$$
\item $(B8\oplus A_1|B7_0\oplus A_1; \operatorname{P2})$

$$[\widetilde{T}^1, \widetilde{T}^4]=- \widetilde{T}^2, \quad [\widetilde{T}^2, \widetilde{T}^4]= \widetilde{T}^1.$$
\end{enumerate}
\subsection{Manin triples with $\mathfrak{g}=B9\oplus A_1$}
Lie products of algebra $\mathfrak{g}=B9\oplus A_1$:
$$[T_1, T_2]= T_3, \quad [T_1, T_3]=- T_2, \quad [T_2, T_3]= T_1.$$
Manin triples, permutations and Lie products of algebra $\tilde{\mathfrak{g}}$:
\begin{enumerate}
\item $(B9\oplus A_1|B5\oplus A_1; \operatorname{P1}, \gamma\,)$

$$[\widetilde{T}^1, \widetilde{T}^2]=-\gamma \widetilde{T}^2, \quad [\widetilde{T}^1, \widetilde{T}^3]=-\gamma \widetilde{T}^3.$$
\item $(B9\oplus A_1|B7_0\oplus A_1; \operatorname{P2})$

$$[\widetilde{T}^1, \widetilde{T}^4]=- \widetilde{T}^2, \quad [\widetilde{T}^2, \widetilde{T}^4]= \widetilde{T}^1.$$
\end{enumerate}
\subsection{Manin triples with $\mathfrak{g}=n_{4,1}$}
Lie products of algebra $\mathfrak{g}=n_{4,1}$:
$$[T_2, T_4]= T_1, \quad [T_3, T_4]= T_2.$$
Manin triples, permutations and Lie products of algebra $\tilde{\mathfrak{g}}$:
\begin{enumerate}
\item $(n_{4,1}|B2\oplus A_1; \operatorname{P9})$

$$[\widetilde{T}^1, \widetilde{T}^2]= \widetilde{T}^3.$$
\item $(n_{4,1}|B2\oplus A_1; \operatorname{P10}, \epsilon\,)$

$$[\widetilde{T}^1, \widetilde{T}^2]=\epsilon \widetilde{T}^4.$$
\item $(n_{4,1}|B2\oplus A_1; \operatorname{P11})$

$$[\widetilde{T}^1, \widetilde{T}^4]= \widetilde{T}^3.$$
\item $(n_{4,1}|B2\oplus A_1; \operatorname{P12})$

$$[\widetilde{T}^1, \widetilde{T}^3]= \widetilde{T}^4.$$
\item $(n_{4,1}|B2\oplus A_1; \operatorname{P22}, \epsilon\,)$

$$[\widetilde{T}^2, \widetilde{T}^3]=\epsilon \widetilde{T}^4.$$
\item $(n_{4,1}|B4\oplus A_1; \operatorname{P1})$

$$[\widetilde{T}^1, \widetilde{T}^2]=- \widetilde{T}^2+ \widetilde{T}^3, \quad [\widetilde{T}^1, \widetilde{T}^3]=- \widetilde{T}^3.$$
\item $(n_{4,1}|B4\oplus A_1; \operatorname{P2}, \epsilon\,)$

$$[\widetilde{T}^1, \widetilde{T}^2]=-\epsilon \widetilde{T}^2+\epsilon \widetilde{T}^4, \quad [\widetilde{T}^1, \widetilde{T}^4]=-\epsilon \widetilde{T}^4.$$
\item $(n_{4,1}|B4\oplus A_1; \operatorname{P8}, \beta\,)$

$$[\widetilde{T}^1, \widetilde{T}^2]=\beta \widetilde{T}^1-\beta \widetilde{T}^4, \quad [\widetilde{T}^2, \widetilde{T}^4]=-\beta \widetilde{T}^4.$$
\item $(n_{4,1}|B5\oplus A_1; \operatorname{P1})$

$$[\widetilde{T}^1, \widetilde{T}^2]=- \widetilde{T}^2, \quad [\widetilde{T}^1, \widetilde{T}^3]=- \widetilde{T}^3.$$
\item $(n_{4,1}|B5\oplus A_1; \operatorname{P2})$

$$[\widetilde{T}^1, \widetilde{T}^2]=- \widetilde{T}^2, \quad [\widetilde{T}^1, \widetilde{T}^4]=- \widetilde{T}^4.$$
\item $(n_{4,1}|B5\oplus A_1; \operatorname{P8})$

$$[\widetilde{T}^1, \widetilde{T}^2]= \widetilde{T}^1, \quad [\widetilde{T}^2, \widetilde{T}^4]=- \widetilde{T}^4.$$
\item $(n_{4,1}|B6_0\oplus A_1; \operatorname{P17})$

$$[\widetilde{T}^1, \widetilde{T}^3]=- \widetilde{T}^4, \quad [\widetilde{T}^1, \widetilde{T}^4]=- \widetilde{T}^3.$$
\item $(n_{4,1}|B7_0\oplus A_1; \operatorname{P17})$

$$[\widetilde{T}^1, \widetilde{T}^3]= \widetilde{T}^4, \quad [\widetilde{T}^1, \widetilde{T}^4]=- \widetilde{T}^3.$$
\item $(n_{4,1}|n_{4,1}; \operatorname{P18}, \epsilon\,)$

$$[\widetilde{T}^1, \widetilde{T}^2]=\epsilon \widetilde{T}^3, \quad [\widetilde{T}^2, \widetilde{T}^3]=-\epsilon \widetilde{T}^4.$$
\item $(n_{4,1}|n_{4,1}; \operatorname{P23}, \epsilon\,)$

$$[\widetilde{T}^1, \widetilde{T}^2]=-\epsilon \widetilde{T}^4, \quad [\widetilde{T}^1, \widetilde{T}^4]=-\epsilon \widetilde{T}^3.$$
\item $(n_{4,1}|n_{4,1}; \operatorname{P24})$

$$[\widetilde{T}^1, \widetilde{T}^2]=- \widetilde{T}^3, \quad [\widetilde{T}^1, \widetilde{T}^3]=- \widetilde{T}^4.$$
\end{enumerate}
\subsection{Manin triples with $\mathfrak{g}=s_{4,1}$}
Lie products of algebra $\mathfrak{g}=s_{4,1}$:
$$[T_2, T_4]=- T_1, \quad [T_3, T_4]=- T_3.$$
Manin triples, permutations and Lie products of algebra $\tilde{\mathfrak{g}}$:
\begin{enumerate}
\item $(s_{4,1}|B2\oplus A_1; \operatorname{P10}, \epsilon\,)$

$$[\widetilde{T}^1, \widetilde{T}^2]=\epsilon \widetilde{T}^4.$$
\item $(s_{4,1}|B2\oplus A_1; \operatorname{P12})$

$$[\widetilde{T}^1, \widetilde{T}^3]= \widetilde{T}^4.$$
\item $(s_{4,1}|B2\oplus A_1; \operatorname{P22})$

$$[\widetilde{T}^2, \widetilde{T}^3]= \widetilde{T}^4.$$
\item $(s_{4,1}|s_{2,1}\oplus A_2; \operatorname{P1})$

$$[\widetilde{T}^1, \widetilde{T}^2]= \widetilde{T}^2.$$
\item $(s_{4,1}|s_{2,1}\oplus A_2; \operatorname{P3})$

$$[\widetilde{T}^1, \widetilde{T}^3]= \widetilde{T}^3.$$
\item $(s_{4,1}|B5\oplus A_1; \operatorname{P1})$

$$[\widetilde{T}^1, \widetilde{T}^2]=- \widetilde{T}^2, \quad [\widetilde{T}^1, \widetilde{T}^3]=- \widetilde{T}^3.$$
\item $(s_{4,1}|s_{4,1}; \operatorname{P22}, \beta\,)$

$$[\widetilde{T}^1, \widetilde{T}^2]=\beta \widetilde{T}^4, \quad [\widetilde{T}^1, \widetilde{T}^3]=\beta \widetilde{T}^3.$$
\item $(s_{4,1}|s_{4,1}; \operatorname{P24})$

$$[\widetilde{T}^1, \widetilde{T}^2]= \widetilde{T}^2, \quad [\widetilde{T}^1, \widetilde{T}^3]= \widetilde{T}^4.$$
\end{enumerate}
\subsection{Manin triples with $\mathfrak{g}=s_{4,2}$}
Lie products of algebra $\mathfrak{g}=s_{4,2}$:
$$[T_1, T_4]=- T_1, \quad [T_2, T_4]=- T_1- T_2, \quad [T_3, T_4]=- T_2- T_3.$$
Manin triples, permutations and Lie products of algebra $\tilde{\mathfrak{g}}$:
\begin{enumerate}
\item $(s_{4,2}|B2\oplus A_1; \operatorname{P10}, \epsilon\,)$

$$[\widetilde{T}^1, \widetilde{T}^2]=\epsilon \widetilde{T}^4.$$
\item $(s_{4,2}|B2\oplus A_1; \operatorname{P12}, \epsilon\,)$

$$[\widetilde{T}^1, \widetilde{T}^3]=\epsilon \widetilde{T}^4.$$
\item $(s_{4,2}|B2\oplus A_1; \operatorname{P22}, \epsilon\,)$

$$[\widetilde{T}^2, \widetilde{T}^3]=\epsilon \widetilde{T}^4.$$
\end{enumerate}
\subsection{Manin triples with $\mathfrak{g}=s_{4,6}$}\label{s46}
Lie products of algebra $\mathfrak{g}=s_{4,6}$:
$$[T_2, T_3]= T_1, \quad [T_2, T_4]=- T_2, \quad [T_3, T_4]= T_3.$$
Manin triples, permutations and Lie products of algebra $\tilde{\mathfrak{g}}$:
\begin{enumerate}
\item $(s_{4,6}|B2\oplus A_1; \operatorname{P10})$

$$[\widetilde{T}^1, \widetilde{T}^2]= \widetilde{T}^4.$$
\item $(s_{4,6}|s_{2,1}\oplus A_2; \operatorname{P1})$

$$[\widetilde{T}^1, \widetilde{T}^2]= \widetilde{T}^2.$$
\item $(s_{4,6}|B5\oplus A_1; \operatorname{P1})$

$$[\widetilde{T}^1, \widetilde{T}^2]=- \widetilde{T}^2, \quad [\widetilde{T}^1, \widetilde{T}^3]=- \widetilde{T}^3.$$
\item $(s_{4,6}|s_{4,1}; \operatorname{P22})$

$$[\widetilde{T}^1, \widetilde{T}^2]= \widetilde{T}^4, \quad [\widetilde{T}^1, \widetilde{T}^3]= \widetilde{T}^3.$$
\end{enumerate}
\subsection{Manin triples with $\mathfrak{g}=s_{4,7}$}\label{s47}
Lie products of algebra $\mathfrak{g}=s_{4,7}$:
$$[T_2, T_3]= T_1, \quad [T_2, T_4]= T_3, \quad [T_3, T_4]=- T_2.$$
Manin triples, permutations and Lie products of algebra $\tilde{\mathfrak{g}}$:
\begin{enumerate}
\item $(s_{4,7}|B2\oplus A_1; \operatorname{P10})$

$$[\widetilde{T}^1, \widetilde{T}^2]= \widetilde{T}^4.$$
\item $(s_{4,7}|B5\oplus A_1; \operatorname{P1})$

$$[\widetilde{T}^1, \widetilde{T}^2]=- \widetilde{T}^2, \quad [\widetilde{T}^1, \widetilde{T}^3]=- \widetilde{T}^3.$$
\item $(s_{4,7}|B7_0\oplus A_1; \operatorname{P13}, \epsilon\,)$

$$[\widetilde{T}^1, \widetilde{T}^2]=\epsilon \widetilde{T}^3, \quad [\widetilde{T}^1, \widetilde{T}^3]=-\epsilon \widetilde{T}^2.$$
\end{enumerate}
\subsection{Manin triples with $\mathfrak{g}=s_{4,10}$}
Lie products of algebra $\mathfrak{g}=s_{4,10}$:
$$[T_1, T_4]=-2 T_1, \quad [T_2, T_3]= T_1, \quad [T_2, T_4]=- T_2, \quad [T_3, T_4]=- T_2- T_3.$$
Manin triples, permutations and Lie products of algebra $\tilde{\mathfrak{g}}$:
\begin{enumerate}
\item $(s_{4,10}|B2\oplus A_1; \operatorname{P10})$

$$[\widetilde{T}^1, \widetilde{T}^2]= \widetilde{T}^4.$$
\item $(s_{4,10}|B2\oplus A_1; \operatorname{P12})$

$$[\widetilde{T}^1, \widetilde{T}^3]= \widetilde{T}^4.$$
\end{enumerate}
\subsection{Manin triples with $\mathfrak{g}=s_{4,11}$}
Lie products of algebra $\mathfrak{g}=s_{4,11}$:
$$[T_1, T_4]=- T_1, \quad [T_2, T_3]= T_1, \quad [T_2, T_4]=- T_2.$$
Manin triples, permutations and Lie products of algebra $\tilde{\mathfrak{g}}$:
\begin{enumerate}
\item $(s_{4,11}|B2\oplus A_1; \operatorname{P7}, \epsilon\,)$

$$[\widetilde{T}^1, \widetilde{T}^3]=\epsilon \widetilde{T}^2.$$
\item $(s_{4,11}|B2\oplus A_1; \operatorname{P8})$

$$[\widetilde{T}^1, \widetilde{T}^4]= \widetilde{T}^2.$$
\item $(s_{4,11}|B2\oplus A_1; \operatorname{P10})$

$$[\widetilde{T}^1, \widetilde{T}^2]= \widetilde{T}^4.$$
\item $(s_{4,11}|B2\oplus A_1; \operatorname{P12})$

$$[\widetilde{T}^1, \widetilde{T}^3]= \widetilde{T}^4.$$
\item $(s_{4,11}|B4\oplus A_1; \operatorname{P9}, \beta\,)$

$$[\widetilde{T}^1, \widetilde{T}^3]=\beta \widetilde{T}^1-\beta \widetilde{T}^2, \quad [\widetilde{T}^2, \widetilde{T}^3]=\beta \widetilde{T}^2.$$
\item $(s_{4,11}|B5\oplus A_1; \operatorname{P9})$

$$[\widetilde{T}^1, \widetilde{T}^3]= \widetilde{T}^1, \quad [\widetilde{T}^2, \widetilde{T}^3]= \widetilde{T}^2.$$
\item $(s_{4,11}|B6_0\oplus A_1; \operatorname{P14})$

$$[\widetilde{T}^1, \widetilde{T}^2]=- \widetilde{T}^4, \quad [\widetilde{T}^1, \widetilde{T}^4]=- \widetilde{T}^2.$$
\item $(s_{4,11}|B7_0\oplus A_1; \operatorname{P14})$

$$[\widetilde{T}^1, \widetilde{T}^2]= \widetilde{T}^4, \quad [\widetilde{T}^1, \widetilde{T}^4]=- \widetilde{T}^2.$$
\item $(s_{4,11}|n_{4,1}; \operatorname{P20})$

$$[\widetilde{T}^1, \widetilde{T}^3]=- \widetilde{T}^4, \quad [\widetilde{T}^1, \widetilde{T}^4]=- \widetilde{T}^2.$$
\item $(s_{4,11}|n_{4,1}; \operatorname{P22}, \epsilon\,)$

$$[\widetilde{T}^1, \widetilde{T}^2]=-\epsilon \widetilde{T}^4, \quad [\widetilde{T}^1, \widetilde{T}^3]=-\epsilon \widetilde{T}^2.$$
\item $(s_{4,11}|s_{4,6}; \operatorname{P11})$

$$[\widetilde{T}^1, \widetilde{T}^2]=- \widetilde{T}^1, \quad [\widetilde{T}^1, \widetilde{T}^4]= \widetilde{T}^3, \quad [\widetilde{T}^2, \widetilde{T}^4]=- \widetilde{T}^4.$$
\item $(s_{4,11}|s_{4,11}; \operatorname{P8})$

$$[\widetilde{T}^1, \widetilde{T}^3]=- \widetilde{T}^1, \quad [\widetilde{T}^1, \widetilde{T}^4]= \widetilde{T}^2, \quad [\widetilde{T}^2, \widetilde{T}^3]=- \widetilde{T}^2.$$
\item $(s_{4,11}|s_{4,11}; \operatorname{P24})$

$$[\widetilde{T}^1, \widetilde{T}^3]= \widetilde{T}^3, \quad [\widetilde{T}^1, \widetilde{T}^4]= \widetilde{T}^4, \quad [\widetilde{T}^2, \widetilde{T}^3]=- \widetilde{T}^4.$$
\end{enumerate}
\subsection{Manin triples with $\mathfrak{g}=s_{4,12}$}
Lie products of algebra $\mathfrak{g}=s_{4,12}$:
$$[T_1, T_3]=- T_1, \quad [T_1, T_4]= T_2, \quad [T_2, T_3]=- T_2, \quad [T_2, T_4]=- T_1.$$
Manin triples, permutations and Lie products of algebra $\tilde{\mathfrak{g}}$:
\begin{enumerate}
\item $(s_{4,12}|B2\oplus A_1; \operatorname{P9})$

$$[\widetilde{T}^1, \widetilde{T}^2]= \widetilde{T}^3.$$
\item $(s_{4,12}|B5\oplus A_1; \operatorname{P10}, \gamma\,)$

$$[\widetilde{T}^1, \widetilde{T}^4]=\gamma \widetilde{T}^1, \quad [\widetilde{T}^2, \widetilde{T}^4]=\gamma \widetilde{T}^2.$$
\item $(s_{4,12}|B7_0\oplus A_1; \operatorname{P1}, \gamma\,)$

$$[\widetilde{T}^1, \widetilde{T}^3]=-\gamma \widetilde{T}^2, \quad [\widetilde{T}^2, \widetilde{T}^3]=\gamma \widetilde{T}^1.$$
\item $(s_{4,12}|B8\oplus A_1; \operatorname{P1}, \gamma\,)$

$$[\widetilde{T}^1, \widetilde{T}^2]=-\gamma \widetilde{T}^3, \quad [\widetilde{T}^1, \widetilde{T}^3]=-\gamma \widetilde{T}^2, \quad [\widetilde{T}^2, \widetilde{T}^3]=\gamma \widetilde{T}^1.$$
\item $(s_{4,12}|B9\oplus A_1; \operatorname{P1}, \gamma\,)$

$$[\widetilde{T}^1, \widetilde{T}^2]=\gamma \widetilde{T}^3, \quad [\widetilde{T}^1, \widetilde{T}^3]=-\gamma \widetilde{T}^2, \quad [\widetilde{T}^2, \widetilde{T}^3]=\gamma \widetilde{T}^1.$$
\item $(s_{4,12}|s_{4,6}; \operatorname{P22})$

$$[\widetilde{T}^1, \widetilde{T}^2]= \widetilde{T}^2, \quad [\widetilde{T}^1, \widetilde{T}^3]=- \widetilde{T}^3, \quad [\widetilde{T}^2, \widetilde{T}^3]= \widetilde{T}^4.$$
\item $(s_{4,12}|s_{4,12}; \operatorname{P1}, \beta\,)$

$$[\widetilde{T}^1, \widetilde{T}^3]=-\beta \widetilde{T}^1, \quad [\widetilde{T}^1, \widetilde{T}^4]=\beta \widetilde{T}^2, \quad [\widetilde{T}^2, \widetilde{T}^3]=-\beta \widetilde{T}^2, \quad [\widetilde{T}^2, \widetilde{T}^4]=-\beta \widetilde{T}^1.$$
\item $(s_{4,12}|s_{4,12}; \operatorname{P2}, \gamma\,)$

$$[\widetilde{T}^1, \widetilde{T}^3]=\gamma \widetilde{T}^2, \quad [\widetilde{T}^1, \widetilde{T}^4]=-\gamma \widetilde{T}^1, \quad [\widetilde{T}^2, \widetilde{T}^3]=-\gamma \widetilde{T}^1, \quad [\widetilde{T}^2, \widetilde{T}^4]=-\gamma \widetilde{T}^2.$$
\item $(s_{4,12}|s_{4,12}; \operatorname{P8}, \gamma\,)$

$$[\widetilde{T}^1, \widetilde{T}^3]=-\gamma \widetilde{T}^2, \quad [\widetilde{T}^1, \widetilde{T}^4]=-\gamma \widetilde{T}^1, \quad [\widetilde{T}^2, \widetilde{T}^3]=\gamma \widetilde{T}^1, \quad [\widetilde{T}^2, \widetilde{T}^4]=-\gamma \widetilde{T}^2.$$
\item $(s_{4,12}|s_{4,12}; \operatorname{P17})$

$$[\widetilde{T}^1, \widetilde{T}^3]= \widetilde{T}^3, \quad [\widetilde{T}^1, \widetilde{T}^4]= \widetilde{T}^4, \quad [\widetilde{T}^2, \widetilde{T}^3]=- \widetilde{T}^4, \quad [\widetilde{T}^2, \widetilde{T}^4]= \widetilde{T}^3.$$
\end{enumerate}
\section{Manin triples for algebras  with one parameter}\label{sec:one_param}

\subsection{Manin triples with $\mathfrak{g}=B6_a\oplus A_1$}
Lie products of algebra $\mathfrak{g}=B6_a\oplus A_1$:
$$[T_1, T_2]=-a T_2- T_3, \quad [T_1, T_3]=- T_2-a T_3,\quad a>0,\ a\neq 1.$$
Manin triples, permutations and Lie products of algebra $\tilde{\mathfrak{g}}$:
\begin{enumerate}
\item $(B6_a\oplus A_1; a|B2\oplus A_1; \operatorname{P1})$

$$[\widetilde{T}^2, \widetilde{T}^3]= \widetilde{T}^1.$$
\item $(B6_a\oplus A_1; a|B2\oplus A_1; \operatorname{P2})$

$$[\widetilde{T}^2, \widetilde{T}^4]= \widetilde{T}^1.$$
\item $(B6_a\oplus A_1; a|B5\oplus A_1; \operatorname{P22})$

$$[\widetilde{T}^2, \widetilde{T}^4]= \widetilde{T}^2, \quad [\widetilde{T}^3, \widetilde{T}^4]= \widetilde{T}^3.$$
\item $(B6_a\oplus A_1; a|B6_0\oplus A_1; \operatorname{P19})$

$$[\widetilde{T}^2, \widetilde{T}^4]= \widetilde{T}^3, \quad [\widetilde{T}^3, \widetilde{T}^4]= \widetilde{T}^2.$$
\item $(B6_a\oplus A_1; a|B6_{a'}\oplus A_1; a', \operatorname{P22}), \quad a'\in \real$

$$[\widetilde{T}^2, \widetilde{T}^4]=a' \widetilde{T}^2+ \widetilde{T}^3, \quad [\widetilde{T}^3, \widetilde{T}^4]= \widetilde{T}^2+a' \widetilde{T}^3.$$
\item $(B6_a\oplus A_1; a|B6_{a'}\oplus A_1; a'=\frac{1}{a}, \operatorname{P1}, \beta\,)$

$$[\widetilde{T}^1, \widetilde{T}^2]=-\frac{\beta }{a} \widetilde{T}^2-\beta \widetilde{T}^3, \quad [\widetilde{T}^1, \widetilde{T}^3]=-\beta \widetilde{T}^2-\frac{\beta }{a} \widetilde{T}^3.$$
\end{enumerate}
\subsection{Manin triples with $\mathfrak{g}=B7_a\oplus A_1$}
Lie products of algebra $\mathfrak{g}=B7_a\oplus A_1$:
$$[T_1, T_2]=-a T_2+ T_3, \quad [T_1, T_3]=- T_2-a T_3,\quad a>0.$$
Manin triples, permutations and Lie products of algebra $\tilde{\mathfrak{g}}$:
\begin{enumerate}
\item $(B7_a\oplus A_1; a|B2\oplus A_1; \operatorname{P1}, \epsilon\,)$

$$[\widetilde{T}^2, \widetilde{T}^3]=\epsilon \widetilde{T}^1.$$
\item $(B7_a\oplus A_1; a|B2\oplus A_1; \operatorname{P2})$

$$[\widetilde{T}^2, \widetilde{T}^4]= \widetilde{T}^1.$$
\item $(B7_a\oplus A_1; a|B5\oplus A_1; \operatorname{P22})$

$$[\widetilde{T}^2, \widetilde{T}^4]= \widetilde{T}^2, \quad [\widetilde{T}^3, \widetilde{T}^4]= \widetilde{T}^3.$$
\item $(B7_a\oplus A_1; a|B7_0\oplus A_1; \operatorname{P19})$

$$[\widetilde{T}^2, \widetilde{T}^4]=- \widetilde{T}^3, \quad [\widetilde{T}^3, \widetilde{T}^4]= \widetilde{T}^2.$$
\item $(B7_a\oplus A_1; a|s_{4,7}; \operatorname{P1}, \epsilon\,)$

$$[\widetilde{T}^2, \widetilde{T}^3]=\epsilon \widetilde{T}^1, \quad [\widetilde{T}^2, \widetilde{T}^4]=\epsilon \widetilde{T}^3, \quad [\widetilde{T}^3, \widetilde{T}^4]=-\epsilon \widetilde{T}^2.$$
\item $(B7_a\oplus A_1; a|B7_{a'}\oplus A_1; a', \operatorname{P22}), \quad a'\in \real$

$$[\widetilde{T}^2, \widetilde{T}^4]=a' \widetilde{T}^2- \widetilde{T}^3, \quad [\widetilde{T}^3, \widetilde{T}^4]= \widetilde{T}^2+a' \widetilde{T}^3.$$
\item $(B7_a\oplus A_1; a|B7_{a'}\oplus A_1; a'=\frac{1}{a}, \operatorname{P1}, \beta\,)$

$$[\widetilde{T}^1, \widetilde{T}^2]=-\frac{\beta }{a} \widetilde{T}^2+\beta \widetilde{T}^3, \quad [\widetilde{T}^1, \widetilde{T}^3]=-\beta \widetilde{T}^2-\frac{\beta }{a} \widetilde{T}^3.$$
\end{enumerate}
\subsection{Manin triples with $\mathfrak{g}=s^{a}_{4,4}$}
Lie products of algebra $\mathfrak{g}=s^{a}_{4,4}$:
$$[T_1, T_4]=- T_1, \quad [T_2, T_4]=- T_1- T_2, \quad [T_3, T_4]=-a T_3, \quad a\neq 0.$$
Manin triples, permutations and Lie products of algebra $\tilde{\mathfrak{g}}$:
\begin{enumerate}
\item $(s^{a}_{4,4}; a|B2\oplus A_1; \operatorname{P10}, \epsilon\,)$

$$[\widetilde{T}^1, \widetilde{T}^2]=\epsilon \widetilde{T}^4.$$
\item $(s^{a}_{4,4}; a|B2\oplus A_1; \operatorname{P12})$

$$[\widetilde{T}^1, \widetilde{T}^3]= \widetilde{T}^4.$$
\item $(s^{a}_{4,4}; a|B2\oplus A_1; \operatorname{P22})$

$$[\widetilde{T}^2, \widetilde{T}^3]= \widetilde{T}^4.$$
\item $(s^{a}_{4,4}; a=2|B2\oplus A_1; \operatorname{P9})$

$$[\widetilde{T}^1, \widetilde{T}^2]= \widetilde{T}^3.$$
\item $(s^{a}_{4,4}; a=2|n_{4,1}; \operatorname{P18})$

$$[\widetilde{T}^1, \widetilde{T}^2]= \widetilde{T}^3, \quad [\widetilde{T}^2, \widetilde{T}^3]=- \widetilde{T}^4.$$
\item $(s^{a}_{4,4}; a=2|n_{4,1}; \operatorname{P24})$

$$[\widetilde{T}^1, \widetilde{T}^2]=- \widetilde{T}^3, \quad [\widetilde{T}^1, \widetilde{T}^3]=- \widetilde{T}^4.$$
\item $(s^{a}_{4,4}; a=-1|s_{4,2}; \operatorname{P16})$

$$[\widetilde{T}^1, \widetilde{T}^3]=- \widetilde{T}^1- \widetilde{T}^2, \quad [\widetilde{T}^2, \widetilde{T}^3]=- \widetilde{T}^2- \widetilde{T}^4, \quad [\widetilde{T}^3, \widetilde{T}^4]= \widetilde{T}^4.$$
\item $(s^{a}_{4,4}; a=-2|s_{4,10}; \operatorname{P16}, \epsilon\,)$

$$[\widetilde{T}^1, \widetilde{T}^2]=-\epsilon \widetilde{T}^4, \quad [\widetilde{T}^1, \widetilde{T}^3]=-\epsilon \widetilde{T}^1-\epsilon \widetilde{T}^2, \quad [\widetilde{T}^2, \widetilde{T}^3]=-\epsilon \widetilde{T}^2, \quad [\widetilde{T}^3, \widetilde{T}^4]=2 \epsilon \widetilde{T}^4.$$
\item $(s^{a}_{4,4}; a|s^{a'}_{4,4}; a'=-a, \operatorname{P8})$

$$[\widetilde{T}^1, \widetilde{T}^3]=- \widetilde{T}^1- \widetilde{T}^2, \quad [\widetilde{T}^2, \widetilde{T}^3]=- \widetilde{T}^2, \quad [\widetilde{T}^3, \widetilde{T}^4]=-a \widetilde{T}^4.$$
\item $(s^{a}_{4,4}; a=-1|s^{a'}_{4,4}; a'=-1, \operatorname{P24})$

$$[\widetilde{T}^1, \widetilde{T}^2]=- \widetilde{T}^2, \quad [\widetilde{T}^1, \widetilde{T}^3]= \widetilde{T}^3+ \widetilde{T}^4, \quad [\widetilde{T}^1, \widetilde{T}^4]= \widetilde{T}^4.$$
\end{enumerate}
\subsection{Manin triples with $\mathfrak{g}=s^{a}_{4,8}$}
Lie products of algebra $\mathfrak{g}=s^{a}_{4,8}$:
$$[T_1, T_4]=-(a+1) T_1, \quad [T_2, T_3]= T_1, \quad [T_2, T_4]=- T_2, \quad [T_3, T_4]=-a T_3,$$
where $-1 < a \leq 1, \ a \neq 0$.

Manin triples, permutations and Lie products of algebra $\tilde{\mathfrak{g}}$:
\begin{enumerate}
\item $(s^{a}_{4,8}; a|B2\oplus A_1; \operatorname{P10})$

$$[\widetilde{T}^1, \widetilde{T}^2]= \widetilde{T}^4.$$
\item $(s^{a}_{4,8}; a|B2\oplus A_1; \operatorname{P12}), \ a \neq 1$

$$[\widetilde{T}^1, \widetilde{T}^3]= \widetilde{T}^4.$$
\item $(s^{a}_{4,8}; a=-\frac{1}{2}|s^{a'}_{4,4}; a'=-2, \operatorname{P24})$

$$[\widetilde{T}^1, \widetilde{T}^2]=-2 \widetilde{T}^2, \quad [\widetilde{T}^1, \widetilde{T}^3]= \widetilde{T}^3+ \widetilde{T}^4, \quad [\widetilde{T}^1, \widetilde{T}^4]= \widetilde{T}^4.$$
\item $(s^{a}_{4,8}; a|s^{a'}_{4,8}; a'=-a-1, \operatorname{P17})$

$$[\widetilde{T}^1, \widetilde{T}^2]=(a+1) \widetilde{T}^1, \quad [\widetilde{T}^1, \widetilde{T}^4]=- \widetilde{T}^3, \quad [\widetilde{T}^2, \widetilde{T}^3]=-a \widetilde{T}^3, \quad [\widetilde{T}^2, \widetilde{T}^4]= \widetilde{T}^4.$$
\item $(s^{a}_{4,8}; a=\frac{1}{2} \left(\sqrt{5}-3\right)|s^{a'}_{4,8}; a'=-\frac{1}{2} \left(3+\sqrt{5}\right), \operatorname{P22})$

\begin{align*}
[\widetilde{T}^1, \widetilde{T}^2] &= \widetilde{T}^2, &[\widetilde{T}^1, \widetilde{T}^3]&=-\frac{1}{2} \left(3+\sqrt{5}\right) \widetilde{T}^3, \\ [\widetilde{T}^1, \widetilde{T}^4] &=-\frac{1}{2} \left(1+\sqrt{5}\right) \widetilde{T}^4, &[\widetilde{T}^2, \widetilde{T}^3]&= \widetilde{T}^4.
\end{align*}
\end{enumerate}
\subsection{Manin triples with $\mathfrak{g}=s^{a}_{4,9}$}
Lie products of algebra $\mathfrak{g}=s^{a}_{4,9}$:
$$[T_1, T_4]=-2 a T_1, \quad [T_2, T_3]= T_1, \quad [T_2, T_4]=-a T_2+ T_3, \quad [T_3, T_4]=- T_2-a T_3$$
where $a > 0$.

Manin triples, permutations and Lie products of algebra $\tilde{\mathfrak{g}}$:
\begin{enumerate}
\item $(s^{a}_{4,9}; a|B2\oplus A_1; \operatorname{P10})$

$$[\widetilde{T}^1, \widetilde{T}^2]= \widetilde{T}^4.$$
\end{enumerate}

\section{Manin triples for algebras  with two parameters}\label{sec:two_param}

\subsection{{Manin triples with $\mathfrak{g}=s^{ab}_{4,3}$}}
Lie products of algebra $\mathfrak{g}=s^{ab}_{4,3}$:
$$[T_1, T_4]=- T_1, \quad [T_2, T_4]=-a T_2, \quad [T_3, T_4]=-b T_3$$
where $(b=-1 \land 0<a\leq 1) \lor (-1<b\leq a\leq 1)$.

Manin triples, permutations and Lie products of algebra $\tilde{\mathfrak{g}}$:
\begin{enumerate}
\item $(s^{ab}_{4,3}; a, b|B2\oplus A_1; \operatorname{P10})$

$$[\widetilde{T}^1, \widetilde{T}^2]= \widetilde{T}^4.$$
\item $(s^{ab}_{4,3}; a, b|B2\oplus A_1; \operatorname{P12}), \quad a \neq b$

$$[\widetilde{T}^1, \widetilde{T}^3]= \widetilde{T}^4.$$
\item $(s^{ab}_{4,3}; a, b|B2\oplus A_1; \operatorname{P22}), \quad a \neq 1$

$$[\widetilde{T}^2, \widetilde{T}^3]= \widetilde{T}^4.$$
\item $(s^{ab}_{4,3}; a, b=1-a|B2\oplus A_1; \operatorname{P1})$

$$[\widetilde{T}^2, \widetilde{T}^3]= \widetilde{T}^1.$$
\item $(s^{ab}_{4,3}; a, b=a-1|B2\oplus A_1; \operatorname{P7}), \quad a \neq 1$

$$[\widetilde{T}^1, \widetilde{T}^3]= \widetilde{T}^2.$$
\item $(s^{ab}_{4,3}; a, b=1-a|n_{4,1}; \operatorname{P10})$

$$[\widetilde{T}^1, \widetilde{T}^3]= \widetilde{T}^4, \quad [\widetilde{T}^2, \widetilde{T}^3]= \widetilde{T}^1.$$
\item $(s^{ab}_{4,3}; a, b=1-a|n_{4,1}; \operatorname{P12}), \quad a \neq \frac{1}{2}$

$$[\widetilde{T}^1, \widetilde{T}^2]= \widetilde{T}^4, \quad [\widetilde{T}^2, \widetilde{T}^3]=- \widetilde{T}^1.$$
\item $(s^{ab}_{4,3}; a, b=a-1|n_{4,1}; \operatorname{P16})$

$$[\widetilde{T}^1, \widetilde{T}^3]= \widetilde{T}^2, \quad [\widetilde{T}^2, \widetilde{T}^3]= \widetilde{T}^4.$$
\item $(s^{ab}_{4,3}; a, b=a-1|n_{4,1}; \operatorname{P22})$

$$[\widetilde{T}^1, \widetilde{T}^2]=- \widetilde{T}^4, \quad [\widetilde{T}^1, \widetilde{T}^3]=- \widetilde{T}^2.$$
\item $(s^{ab}_{4,3}; a, b=-1|s^{a'}_{4,4}; a'=a, \operatorname{P10})$

$$[\widetilde{T}^1, \widetilde{T}^3]=- \widetilde{T}^1- \widetilde{T}^4, \quad [\widetilde{T}^2, \widetilde{T}^3]=-a \widetilde{T}^2, \quad [\widetilde{T}^3, \widetilde{T}^4]= \widetilde{T}^4.$$
\item $(s^{ab}_{4,3}; a, b=-a|s^{a'}_{4,4}; a'=\frac{1}{a}, \operatorname{P16}), \quad a \neq 1$

$$[\widetilde{T}^1, \widetilde{T}^3]=-\frac{1}{a} \widetilde{T}^1, \quad [\widetilde{T}^2, \widetilde{T}^3]=- \widetilde{T}^2- \widetilde{T}^4, \quad [\widetilde{T}^3, \widetilde{T}^4]= \widetilde{T}^4.$$
\item $(s^{ab}_{4,3}; a, b=-a|s^{a'}_{4,4}; a'=-\frac{1}{a}, \operatorname{P18})$

$$[\widetilde{T}^1, \widetilde{T}^2]=\frac{1}{a} \widetilde{T}^1, \quad [\widetilde{T}^2, \widetilde{T}^3]= \widetilde{T}^3+ \widetilde{T}^4, \quad [\widetilde{T}^2, \widetilde{T}^4]= \widetilde{T}^4.$$
\item $(s^{ab}_{4,3}; a, b=-1|s^{a'}_{4,4}; a'=-a, \operatorname{P24}), \quad a \neq 1$

$$[\widetilde{T}^1, \widetilde{T}^2]=-a \widetilde{T}^2, \quad [\widetilde{T}^1, \widetilde{T}^3]= \widetilde{T}^3+ \widetilde{T}^4, \quad [\widetilde{T}^1, \widetilde{T}^4]= \widetilde{T}^4.$$
\item $(s^{ab}_{4,3}; a, b=-a-1|s^{a'}_{4,8}; a'=a, \operatorname{P10})$

$$[\widetilde{T}^1, \widetilde{T}^2]= \widetilde{T}^4, \quad [\widetilde{T}^1, \widetilde{T}^3]=- \widetilde{T}^1, \quad [\widetilde{T}^2, \widetilde{T}^3]=-a \widetilde{T}^2, \quad [\widetilde{T}^3, \widetilde{T}^4]=(a+1) \widetilde{T}^4.$$
\item $(s^{ab}_{4,3}; a, b=-a-1|s^{a'}_{4,8}; a'=-a-1, \operatorname{P12})$

$$[\widetilde{T}^1, \widetilde{T}^2]=- \widetilde{T}^1, \quad [\widetilde{T}^1, \widetilde{T}^3]= \widetilde{T}^4, \quad [\widetilde{T}^2, \widetilde{T}^3]=-(a+1) \widetilde{T}^3, \quad [\widetilde{T}^2, \widetilde{T}^4]=-a \widetilde{T}^4.$$
\item $(s^{ab}_{4,3}; a, b=-a-1|s^{a'}_{4,8}; a'=-\frac{a+1}{a}, \operatorname{P22})$

$$[\widetilde{T}^1, \widetilde{T}^2]= \widetilde{T}^2, \quad [\widetilde{T}^1, \widetilde{T}^3]=-\frac{a+1}{a} \widetilde{T}^3, \quad [\widetilde{T}^1, \widetilde{T}^4]=-\frac{1}{a} \widetilde{T}^4, \quad [\widetilde{T}^2, \widetilde{T}^3]= \widetilde{T}^4.$$
\item $(s^{ab}_{4,3}; a, b|s^{a'b'}_{4,3}; a'=a, b'=-b, \operatorname{P2})$

$$[\widetilde{T}^1, \widetilde{T}^3]=- \widetilde{T}^1, \quad [\widetilde{T}^2, \widetilde{T}^3]=-a \widetilde{T}^2, \quad [\widetilde{T}^3, \widetilde{T}^4]=-b \widetilde{T}^4.$$
\item $(s^{ab}_{4,3}; a, b|s^{a'b'}_{4,3}; a'=b, b'=-a, \operatorname{P5}), \quad a \neq b$

$$[\widetilde{T}^1, \widetilde{T}^2]=- \widetilde{T}^1, \quad [\widetilde{T}^2, \widetilde{T}^3]=b \widetilde{T}^3, \quad [\widetilde{T}^2, \widetilde{T}^4]=-a \widetilde{T}^4.$$
\item $(s^{ab}_{4,3}; a, b|s^{a'b'}_{4,3}; a'=\frac{b}{a}, b'=-\frac{1}{a}, \operatorname{P19}), \quad a \neq 1$

$$[\widetilde{T}^1, \widetilde{T}^2]= \widetilde{T}^2, \quad [\widetilde{T}^1, \widetilde{T}^3]=\frac{b}{a} \widetilde{T}^3, \quad [\widetilde{T}^1, \widetilde{T}^4]=-\frac{1}{a} \widetilde{T}^4.$$
\end{enumerate}
\subsection{Manin triples with $\mathfrak{g}=s^{ab}_{4,5}$}
Lie products of algebra $\mathfrak{g}=s^{ab}_{4,5}$:
$$[T_1, T_4]=-a T_1, \quad [T_2, T_4]=-b T_2+ T_3, \quad [T_3, T_4]=- T_2-b T_3, \quad a > 0.$$
Manin triples, permutations and Lie products of algebra $\tilde{\mathfrak{g}}$:
\begin{enumerate}
\item $(s^{ab}_{4,5}; a, b|B2\oplus A_1; \operatorname{P10})$

$$[\widetilde{T}^1, \widetilde{T}^2]= \widetilde{T}^4.$$
\item $(s^{ab}_{4,5}; a, b|B2\oplus A_1; \operatorname{P22}, \epsilon\,)$

$$[\widetilde{T}^2, \widetilde{T}^3]=\epsilon \widetilde{T}^4.$$
\item $(s^{ab}_{4,5}; a, b=\frac{a}{2}|B2\oplus A_1; \operatorname{P1})$

$$[\widetilde{T}^2, \widetilde{T}^3]= \widetilde{T}^1.$$
\item $(s^{ab}_{4,5}; a, b=\frac{a}{2}|n_{4,1}; \operatorname{P10})$

$$[\widetilde{T}^1, \widetilde{T}^3]= \widetilde{T}^4, \quad [\widetilde{T}^2, \widetilde{T}^3]= \widetilde{T}^1.$$
\item $(s^{ab}_{4,5}; a, b=-\frac{a}{2}|s^{a'}_{4,9}; a'=\frac{a}{2}, \operatorname{P22}, \epsilon\,)$

$$[\widetilde{T}^1, \widetilde{T}^2]=\frac{a \epsilon }{2} \widetilde{T}^2-\epsilon \widetilde{T}^3, \quad [\widetilde{T}^1, \widetilde{T}^3]=\epsilon \widetilde{T}^2+\frac{a \epsilon }{2} \widetilde{T}^3, \quad [\widetilde{T}^1, \widetilde{T}^4]=a \epsilon \widetilde{T}^4, \quad [\widetilde{T}^2, \widetilde{T}^3]=\epsilon \widetilde{T}^4.$$
\item $(s^{ab}_{4,5}; a, b|s^{a'b'}_{4,5}; a'=a, b'=-b, \operatorname{P22})$

$$[\widetilde{T}^1, \widetilde{T}^2]=-b \widetilde{T}^2- \widetilde{T}^3, \quad [\widetilde{T}^1, \widetilde{T}^3]= \widetilde{T}^2-b \widetilde{T}^3, \quad [\widetilde{T}^1, \widetilde{T}^4]=a \widetilde{T}^4.$$
\end{enumerate}

\section{Classification of \dd s}\label{sec:dd_class}

Having the list of standard \mt s, we can ask which of them give the same \dd. This is important if we aim to study not only \pl\ T-duality but also plurality. However, the classification of \dd s is much more complicated than the classification of \mt s. In this section we describe how the classification can be done, and give an example of eight-dimensional \dd\ that can be decomposed in several different ways into \mt s presented in Section \ref{sec:no_param}.

Two \mt s belong to the same \dd\ if and only if they have isomorphic algebraic structure and the isomorphism transforms one ad-invariant bilinear form to the other. More explicitly, as mentioned above, we can always choose bases in the \mt s such that the bilinear forms take the canonical form \eqref{dual_bases} and the Lie product is given by \eqref{commutation_on_d}. The \mt s $MT = (\cd, \cg, \tcg)$ and $MT' = (\cd', \cg', \tcg')$ with these special bases 
$$Y_a=(T_1, T_2, T_3, T_4, \widetilde{T}^1, \widetilde{T}^2,\widetilde{T}^3,\widetilde{T}^4), \quad Y'_a=(T'_1, T'_2, T'_3, T'_4, \widetilde{T}'^1, \widetilde{T}'^2, \widetilde{T}'^3, \widetilde{T}'^4)$$
belong to the same \dd\  if and only if there is an invertible $8\times 8$ matrix $C$ such that the linear map given by
$$ Y'_a = {C_a}^b Y_b \label{isodd}$$
transforms\footnote{Note the difference between this transformation and \tfn\ \eqref{iso_A} of bases of the \mt s.} the Lie products of $MT$ into that of $MT'$ and preserves the canonical form of the bilinear form $\langle.\, ,.\rangle$. 
We denote the structure coefficients of the algebras $\cd, \cd'$ as ${F_{ab}}^c,\ {F'_{ab}}^c$ for $a,b,c=1,\ldots,8$, i.e.
$$
[Y_a,Y_b]={F_{ab}}^c Y_c,
$$
and 
$$
\eta=\left(
\begin{array}{cc}
 0 & \unit_4 \\
 \unit_4 & 0 \\
\end{array}
\right)
$$
where $\unit_4$ is the $4\times 4$ unit matrix. The matrix $C$ then has to satisfy the conditions
\begin{equation}
{C_a}^p {C_b}^q \eta_{pq}= \eta_{ab},\qquad  {C_a}^p {C_b}^q {F_{pq}}^r ={ F'_{ab}}^c {C_c}^r. \label{cpodm}
\end{equation}
The first condition states that $C$ is an element of the $O(4,4)$ group.

It is clear that directly checking which of the 188 standard Manin triples correspond to the same \dd\ is an impossible task. That is why we first evaluate invariants of the algebras $\cd$ for all \mt s, and then sort them into smaller subsets. 
Only those \mt s with the same invariants can belong to the same Drinfeld double. The invariants that we have used are:
\begin{itemize}
\item Dimensions of the derived series
$$
\cd^{0} = \cd, \qquad \cd^{k+1} = [ \cd^{k}, \cd^{k} ], \quad k \in \mathbb{N}.
$$ 
\item Dimensions of the lower central series
$$
\cd_{0} = \cd, \qquad \cd_{k+1} = [ \cd_{k}, \cd], \quad k \in \mathbb{N}.
$$
\item Dimension of the algebra of derivations $Der\, \cd\ni \cal A$
$$
 {\cal A} [X, Y]=[{\cal A} X, Y]+[X,{\cal A} Y], \text{ for all } X, Y \in \cd. 
$$
\item Signature of the Killing form $K_{ab} = {F_{ad}}^c {F_{bc}}^d$ (numbers of its positive, zero and negative eigenvalues). 
\end{itemize}

We have also determined other invariants, but they do not lead to a refinement of the partition. Solving \eqref{cpodm} is computationally demanding and if a solution is not found, it is difficult to prove that \mt s with the same invariants do not correspond to the same \dd. In many cases, however, the isomorphisms can be found.

In the following we will be interested in the \dd\ with subalgebra $\cg = s_{4,6}$. The \dd\ has dimensions of derived series $\{8,6,2,0\}$, dimensions of lower central series $\{8,6\}$, dimension of $Der\, \cd$ equals $12$, and the signature of the Killing form is $\{2,6,0\}$.
The \mt s of this \dd\ are 
\begin{align}
(s_{2,1}\oplus s_{2,1}|A_4) &\cong (s_{2,1}\oplus s_{2,1}|s_{2,1}\oplus s_{2,1}; \operatorname{P8}) \cong \nonumber\\   (s_{2,1}\oplus A_2|s_{2,1}\oplus s_{2,1}; \operatorname{P7}) &\cong (s_{2,1}\oplus A_2|s_{2,1}\oplus A_2; \operatorname{P17}) \cong \nonumber\\
(B5\oplus A_1|s_{2,1}\oplus A_2;\operatorname{P16}) &\cong (B6_0\oplus A_1|B5\oplus A_1; \operatorname{P10}) \cong \label{dds46}\\
(s_{4,1}|s_{2,1}\oplus A_2; \operatorname{P1}) &\cong  (s_{4,1}|B5\oplus A_1; \operatorname{P1}) \cong \nonumber\\
(s_{4,1}|s_{4,1}; \operatorname{P24}) &\cong  (s_{4,6}|s_{2,1}\oplus A_2; \operatorname{P1}) \cong \nonumber\\ 
(s_{4,6}|B5\oplus A_1; \operatorname{P1}) &\cong  (s_{4,6}|s_{4,1}; \operatorname{P22}) \nonumber
\end{align}
where $\cong$ denotes the \dd\ isomorphism.

\section{Applications to the WZW models}\label{sec:wzw}

In this section we will show that some of the above found \mt s can be used for construction of WZW \sm s on four-dimensional groups $\G$.

\subsection{WZW models}

We denote by $x^\mu$ the coordinates on the group manifold $\G$ and $\sigma^{\alpha}=(\sigma^+ , \sigma^-)$ the light-cone variables on the worldsheet $\Sigma$. The action of the WZW model on the Lie group $\G$ is specified by a non-degenerate ad-invariant symmetric bilinear form $\Omega$ on the Lie algebra $\mathfrak{g}$ and can be expressed as 
\begin{equation}\label{2.10}
S_{_{WZW}}(g) =  \frac{1}{2} \int_{\Sigma} d\sigma^+ d\sigma^-\;{\Omega}_{ij} L^{~i}_{_+}\; L^{~j}_{_-} +\frac{1}{12} \int_{M} d^3 \sigma~ \varepsilon^{ \gamma \alpha \beta}~{\Omega}_{ik} \;{f_{jl}}^{k}~ L^{~i}_{_\gamma} L^{~j}_{_\alpha} L^{~l}_{_\beta}
\end{equation}
where ${f_{jl}}^{k}$ are the structure constants of the Lie algebra $\mathfrak{g}$ of the Lie group $\G$, $\varepsilon^{ \gamma \alpha \beta}$ is the Levi--Civita symbol and $M$ is a three-dimensional manifold with boundary $\partial M = \Sigma$. The mapping $g : \Sigma \mapsto \G$ extends to $M$ arbitrarily. The $L^{~i}_{_\alpha}$'s are defined by
 the components of the left-invariant fields on $\G$ as 
\begin{equation}\nonumber 
g^{-1} \partial_\alpha g = L^{~i}_{_\alpha} T_i=\partial_\alpha x^\mu~L^{~i}_{_\mu}~T_i,
\end{equation}
where $T_i$, $i=1, \ldots, D$ form the basis of the Lie algebra $\mathfrak{g}$.

Alternatively, if the 3-form $H$ with components 
\begin{equation}\label{hmnr}
H_{{\mu \nu \rho}}= \Omega_{ik} {f_{jl}}^{k} L_{_\mu}^{~i}~L_{_\nu}^{~j}~L_{_\rho}^{~l}
\end{equation} 
equals to the strength of an antisymmetric $B$-field, i.e.
\begin{equation}\label{condPLWZW}
(d B)_{\mu\nu\rho}=H_{_{\mu \nu \rho}},
\end{equation}
we can regard the WZW model as the 2-dimensional non-linear sigma model with the action
$$
S=\frac{1}{2}\int \!d\sigma^+ d\sigma^- (G_{{\mu\nu}}\ 
+B_{{\mu\nu}})\partial_{_+}x{^\mu} \partial_{_-}x^{\nu}, \quad G_{\mu\nu}=G_{\nu\mu},\quad B_{\mu\nu}=-B_{\nu\mu}
$$
given by metric $G$ and Kalb--Ramond $B$-field on the manifold $\G$.

\subsection{\PL \sm s}

The \sm\ can be called \PL symmetric if the Lie derivatives of $F_{\mu \nu}=G_{\mu\nu}+B_{\mu\nu}$ with respect to the left-invariant vector fields ${V_a}$ of the group $\G$ satisfy the \cond\ \cite{klise}
\begin{equation}\label{KScond}
({\cal L}_{V_a}F)_{\mu \nu}~=~
 F_{\mu \rho} ~{V_b}^{\rho}\;\tilde
f^{cb}{}_{a}\;{V_c}^{\lambda}\;F_{\lambda \nu}
\end{equation}
for structure constants ${\tilde f}^{cb}{}_{a}$ of some dual Lie algebra ${\mathfrak{\tilde g}}$. The self-consistency of the condition \eqref{KScond} implies that algebras $\cg$ and $\tcg$ have to form a \mt.

The general solution of the equation \eqref{KScond} has the form
\begin{equation}\label{met}
F_{\mu\nu}(x)={e_{\mu}}^{a}(g(x))\, E_{ab}(g(x)) \, {e_{\nu}}^{b}(g(x)),
\end{equation}
where ${e_{\mu}}^{a}(g(x))$ are the components of right-invariant forms $dg g^{-1}$ expressed in coordinates $x^\mu$ on the group $\G$,
\begin{equation}\label{metr}
E(g)=\left(E_{0}^{-1}+\Pi(g)\right)^{-1}, \qquad \Pi(g)=b(g) \cdot a(g)^{-1}
\end{equation}
for some constant invertible matrix $E_0$, and matrices $a(g),b(g)$ given by the adjoint representation of the Lie group $\G$ on the Lie algebra of the Drinfeld double $\cd$ in the mutually dual bases 
\begin{equation}
Ad(g^{-1})^{T}=
\left(
\begin{array}{cc}
a(g) & 0 \\
b(g) & d(g) \\
\end{array}
\right).
\end{equation}

Every \mt\ can be used for the construction of \PL \sm\ and its dual, but only a few of the \PL \sm s are WZW models. It was shown in Ref. \cite{kehagias:wzw} that in four dimensions the non-degenerate symmetric bilinear form $\Omega$ satisfying conditions of the ad-invariance
\begin{equation}\label{ad_omega}
{f_{ij}}^{k} \Omega_{kl} + {f_{il}}^{k} \Omega_{kj} = 0
\end{equation}
exists only for the groups $H_4$ and $E_2^c$ whose Lie algebras are isomorphic to $s_{4,6}$ and $s_{4,7}$. Having \mt s containing these algebras, we may construct the WZW models as \PL models. However, not all \mt s with $\mathfrak{g}=s_{4,6}$ or $\mathfrak{g}=s_{4,7}$ generate WZW models.

\subsubsection{\PL $H_4$ WZW models}

\PL construction of $H_4$ WZW model was given in Ref. \cite{eghbali} using the \mt\ isomorphic to $(s_{4,6}|s_{2,1}\oplus A_2; \operatorname{P1})$. Let us recalculate its form in our notation.

The ad-invariant form $\Omega$ satisfying \eqref{ad_omega} for the algebra $s_{4,6}$ has the components
\begin{eqnarray}\label{OH4}
\Omega_{ij} = \left(
\begin{array}{cccc}
 0 & 0 & 0 & \kappa  \\
 0 & 0 & \kappa  & 0 \\
 0 & \kappa  & 0 & 0 \\
 \kappa  & 0 & 0 & \rho \\
\end{array}
\right),
\end{eqnarray}
where $\rho$ and $\kappa$ are arbitrary constants. Using parametrization of the elements of the corresponding group in the form
\begin{equation}\label{solvpar}
g(x)=e^{x^4 T_{4}} ~ e^{x^3 T_{3}} ~e^{x^2 T_{2}} ~e^{x^1 T_{1}},
\end{equation}
we get components of the left-invariant form 
\begin{equation}
L_{_\mu}{}^i= \left(
\begin{array}{cccc}
 1 & 0 & 0 & 0 \\
 0 & 1 & 0 & 0 \\
 -x^2 & 0 & 1 & 0 \\
 x^2 x^3 & x^2 & -x^3 & 1 \\
\end{array}
\right),
\end{equation}
and the corresponding 3-form $H$ then is 
\begin{equation}
H=\kappa\, dx^2\wedge dx^3 \wedge dx^4. \label{tor_4_6}
\end{equation}

By the standard \PL procedure \cite{klise,LH} for $(s_{4,6}|s_{2,1}\oplus A_2; \operatorname{P1})$ and $E_0=\Omega$ we obtain
\begin{equation}\label{wzw1}
F_{\mu\nu}=G_{\mu\nu}+B_{\mu\nu}=\left(
\begin{array}{cccc}
 0 & 0 & 0 & \kappa  \\
 0 & 0 & \kappa  & -\kappa  x^3 \\
 0 & \kappa  & 0 & -\kappa ^2 x^2 \\
 \kappa  & -\kappa  x^3 & \kappa ^2 x^2 & \rho \\
\end{array}
\right).
\end{equation}
It is easy to check the \cond\ \eqref{KScond}. For $\kappa=-1$ the condition $d B = H$ is satisfied and \eqref{wzw1} represents a WZW model. The vanishing beta function equations \cite{hlape:bianchisugra} are satisfied for $\kappa=\pm 1$ and vanishing dilaton $\Phi = 0$.

This form of the WZW model can be transformed to that presented in Ref. \cite{eghbali},
$$ 
{\cal E}_{\mu\nu}=\left(
\begin{array}{cccc}
 \rho & 0 & -e^x y & -1 \\
 0 & 0 & e^x & 0 \\
 e^x y & e^x & 0 & 0 \\
 -1 & 0 & 0 & 0 \\
\end{array}
\right)
$$
by the \coor\ \tfn 
$$ x^1=v,\quad x^2=y,\quad x^3=e^x u,\quad x^4=x. $$

The \mt\ $(s_{4,6}|s_{2,1}\oplus A_2; \operatorname{P1})$ belongs to the \dd\ \eqref{dds46}, and we can use its various decompositions into \mt s to construct plural sigma models to \eqref{wzw1}. However, only two of the \mt s in that \dd\ generate WZW models, namely $(s_{4,6}|s_{2,1}\oplus A_2; \operatorname{P1})$ and $(s_{4,6}|s_{4,1}; \operatorname{P22})$. They are related by the \tfn\ \eqref{cpodm} with 
$$ C=\left(
\begin{array}{cccccccc}
 1 & 0 & 0 & 0 & 0 & 0 & 0 & 0 \\
 0 & 1 & 0 & 0 & 0 & 0 & 0 & 0 \\
 0 & 0 & 1 & 0 & 0 & 0 & 0 & 0 \\
 0 & 0 & 0 & 1 & 0 & 0 & 0 & 0 \\
 0 & -1 & 0 & -1 & 1 & 0 & 0 & 0 \\
 1 & 0 & 0 & 0 & 0 & 1 & 0 & 0 \\
 0 & 0 & 0 & 0 & 0 & 0 & 1 & 0 \\
 1 & 0 & 0 & 0 & 0 & 0 & 0 & 1 \\
\end{array}
\right).$$
Unfortunately, this transformation cannot be used for construction of the WZW model as the plural matrix 
\begin{equation}
\hat{E}_0=\left(
\begin{array}{cccc}
 0 & 0 & 0 & \frac{\kappa }{\kappa +1} \\
 0 & 0 & \kappa  & 0 \\
 0 & \kappa  & 0 & -\frac{\kappa ^2}{\kappa +1} \\
 \frac{\kappa }{1-\kappa } & 0 & \frac{\kappa ^2}{1-\kappa } & \frac{\rho }{1-\kappa ^2} \\
\end{array}
\right)
\end{equation}
is singular for $\kappa=\pm 1$ and is not of the form \eqref{OH4}. 
Nevertheless, other plural models can be found.

In addition to the WZW model given above, there is another WZW model obtained from the \mt\ $(s_{4,6}|s_{4,1}; \operatorname{P22})$. Namely, for this \mt\ and $E_0=\Omega$  we get
\begin{equation}\label{pluralwzw}
F_{\mu\nu}=G_{\mu\nu}+B_{\mu\nu}= \left(
\begin{array}{cccc}
 0 & 0 & 0 & \kappa  \\
 0 & 0 & \kappa  & -\kappa  (\kappa +1) x^3 \\
 0 & \kappa  & 0 & \kappa ^2 \left(e^{-x^4}-1\right) \\
 \kappa  & (\kappa -1) \kappa  x^3 & \kappa ^2 \left(1-e^{-x^4}\right) & \rho+2 \kappa ^3 \left(e^{-x^4}-1\right) x^3 \\
\end{array}
\right)
\end{equation}
and one can check that both \cond s \eqref{KScond} and \eqref{condPLWZW} are satisfied for $\kappa=1$. It means that  the tensor field \eqref{pluralwzw} yields the WZW \sm\ given by
$$ ds^2= 2\,dx^1dx^4+2\,dx^2dx^3-x^3 dx^2dx^4+\left(\rho+\left(2 e^{-x^4}-2\right) x^3\right)dx^4dx^4,$$
$$B= -x^3\, dx^2\wedge dx^4+(e^{-x^4}-1)dx^3\wedge dx^4.$$
It seems that it is not possible to transform the metric of \eqref{wzw1} to that of \eqref{pluralwzw} by a coordinate \tfn. The vanishing beta function equations are satisfied for vanishing dilaton $\Phi = 0$.

\subsubsection{\PL construction of a modified Nappi--Witten model}

The Nappi--Witten WZW model \cite{NW} was reconstructed by a generalized \PL construction in Ref. \cite{saka:wzw}, and by the \PL construction with spectators in Ref. \cite{egh:NW}. Here we are going to show that by the \PL method \cite{klise,LH} using the \mt\ $(s_{4,7},B7_a\oplus A_1,P1,\epsilon=\pm1),\ a\geq 0$ we get a WZW model that resembles the Nappi--Witten model. 

The ad-invariant form $\Omega$ for the algebra $s_{4,7}$ has the components
\begin{equation}\label{ONP}
\Omega_{ij}=\left(
\begin{array}{cccc}
 0 & 0 & 0 & -\kappa  \\
 0 & \kappa  & 0 & 0 \\
 0 & 0 & \kappa  & 0 \\
 -\kappa  & 0 & 0 & \rho \\
\end{array}
\right).
\end{equation}
The parametrization \eqref{solvpar} yields the left-invariant form  given by
\begin{equation}
L_{_\mu}{}^i=\left(
\begin{array}{cccc}
 1 & 0 & 0 & 0 \\
 0 & 1 & 0 & 0 \\
 -x^2 & 0 & 1 & 0 \\
 \frac{1}{2} \left((x^2)^2+(x^3)^2\right) & x^3 & -x^2 & 1 \\
\end{array}
\right),
\end{equation}
and the 3-form \eqref{hmnr} equals
$$ H=-\kappa \, dx^2\wedge dx^3 \wedge dx^4. $$

By the standard \PL procedure for \mt\ $(s_{4,7},B7_a\oplus A_1,P1,\epsilon=\pm1),\ a\geq 0$ and $E_0=\Omega$ with $\kappa=\epsilon/2 $ we get \PL \sm\ which is a WZW model. Its tensor field is 
\begin{align}
F_{\mu\nu}&=G_{\mu\nu}+B_{\mu\nu}=
\label{wzw2}\\
&\left(
\begin{array}{cccc}
 0 & 0 & 0 & -\frac{\epsilon }{2} \\
 0 & \frac{\epsilon }{2} & 0 & \frac{1}{4} \epsilon \left(3 x^3 - a x^2 \right) \\
 0 & 0 & \frac{\epsilon }{2} & -\frac{1}{4} \epsilon  \left(a x^3+x^2\right) \\
 -\frac{\epsilon }{2} & \frac{1}{4} \epsilon  \left(a x^2+x^3\right) & \frac{1}{4} \epsilon  \left(a
   x^3+x^2\right) & \rho-\frac{1}{8}\epsilon \left(a^2+1\right) \left((x^2)^2+(x^3)^2\right)  \\
\end{array}
\right) \nonumber
\end{align}
and  it can be transformed by the \coor\ \tfn
$$
x^1=(2+\epsilon) x y-2\, v,\qquad x^2=\sqrt{2}\, y,\qquad x^3=\sqrt{2}\, x,\qquad x^4 =u
$$
to the form with
\begin{equation}\label{mtk2}
G_{\mu\nu}=\left(
\begin{array}{cccc}
 \epsilon  & 0 & -\frac{1}{2} y \,(2\,\epsilon +1) & 0 \\
 0 & \epsilon  & -\frac{x}{2} & 0 \\
 -\frac{1}{2} y\, (2\,\epsilon +1) & -\frac{x}{2} & \rho-\frac{1}{4}\,\epsilon \left(a^2+1\right)\left( x^2 + 
   y^2 \right) 
   & \epsilon  \\
 0 & 0 & \epsilon  & 0 \\
\end{array}
\right),
\end{equation}
$$ B_{\mu\nu}=\left(
\begin{array}{cccc}
 0 & 0 & -\frac{1}{2} \epsilon\,  (a x+y) & 0 \\
 0 & 0 & \frac{1}{2} \epsilon\,  (x-a y) & 0 \\
 \frac{1}{2} \epsilon\,  (a x+y) & \frac{1}{2} \epsilon\, (a y-x ) & 0 & 0 \\
 0 & 0 & 0 & 0 \\
\end{array}
\right).$$ 
This resembles the Nappi--Witten model in Ref. \cite{NW}, the difference being that $(G_{NW})_{3,3}= \rho$. It seems that it is not possible to transform the plane-wave metric \eqref{mtk2} to that in \cite{NW} by coordinate transformations. Nevertheless, the strength of the field B $$ H=\epsilon\,dx\wedge dy\wedge du $$
is equal to that of the Nappi--Witten model. The vanishing beta function equations are satisfied for dilaton 
$$
\Phi = c_1+c_2 x^4-\frac{1}{8} \left(a^2+1\right) (x^4)^2, \qquad c_1, c_2 = \text{const.}$$
\PL\ models for the \mt s $(s_{4,7},B2\oplus A_1,P10)$ or $(s_{4,7},B5\oplus A_1,P1)$ do not yield WZW models. Another modification of the Nappi--Witten model with generally non-vanishing scalar curvature  was found in \cite{khalid}.

\section{Conclusion}

We have obtained an extensive list of (4+4)-dimensional \mt s $(\cd, \cg, \tcg)$ that can be used to construct \PL symmetric \sm s, their \PL duals, and (after classification of the corresponding \dd s) also \PL plurals. Due to the enormous complexity of complete classification of these \mt s, we focused on \mt s in ``standard'' form, where the algebra $\cg$ belongs to the list  of the four dimensional algebras presented in Ref. \cite{snowin:kniha}, and the dual algebras $ \tcg$ are obtained from these by permutations and scalings of their bases. The results of the classification of these standard \mt s are given in Sections \ref{sec:no_param}--\ref{sec:two_param}.

Furthermore, we have used the \mt s where $\cg$ is  $s_{4,6}$ or $s_{4,7}$ for the construction of WZW models using the Poisson--Lie procedure. We have found two new WZW models. Namely, the $H_4$ WZW model \eqref{pluralwzw} which is different from that given in Refs. \cite{eghbali,kehagias:wzw} and a modification \eqref{mtk2} of the Nappi--Witten model \cite{NW}. 

\appendix
\section{Appendix}
There are 25 non-isomorphic real four-dimensional Lie algebras that were classified in Refs. \cite{Mubarakzyan,pawiza,snowin:kniha}. For the indecomposable algebras listed in Table \ref{4dimalras} we adopted the notation of Ref. \cite{snowin:kniha}. The decomposable algebras have the form
$$
A_4, s_{2,1}\oplus s_{2,1}, B2 \oplus A_1, \ldots, B9\oplus A_1, B6_a\oplus A_1, B7_a\oplus A_1
$$
where $A_1$ and $A_4$ are the one- and four-dimensional Abelian algebras, $s_{2,1}$ is the two-dimensional algebra given in Table \ref{2dimalras}, and $B_i$ refers to three-dimensional algebras in the Bianchi classification summarized in Table \ref{3dimalras}. We use the Bianchi classification to be able to compare our results with classification of $(3+3)$-dimensional \mt s presented in Ref. \cite{snohla:DD}. The relation between Bianchi classification and the classification given in Ref. \cite{snowin:kniha} is the following:

\begin{align*}
n_{3,1} &= B2, & s_{3,1}^a & \cong B6_0, \quad a=-1, \\
s_{2,1}\oplus A_1 & \cong B3, & s_{3,1}^a &= B6_{\kappa}, \quad \kappa = \frac{1+a}{1-a} \\
s_{3,2} &= B4, & s_{3,1}^a &\cong B5, \quad a = 1,\\
sl(2,\real) &= B8, & s_{3,3}^a &= B7_a, \\
so(3,\real) &= B9. & &
\end{align*}
The ordering of four-dimensional algebras $A_{4,j},\ j=0,\ldots,24$ can be found in Table \ref{ordering}. We list the algebras without parameters first. For completeness we add the numbering of permutations in Table \ref{permutations}.

\begin{table}
\begin{center}
\renewcommand*{\arraystretch}{1.5}
\begin{tabular}{| c || l | l |}
\hline
$\cg$ & nontrivial Lie products & parameters\\
\hline
\hline
$s_{2,1}$ & $[T_1,T_2]=T_2$ & \\
\hline
\end{tabular}
\caption{Non-Abelian two-dimensional real Lie algebras as used in \cite{snohla:DD}.}\label{2dimalras}
\end{center}
\end{table}

\begin{table}
\begin{center}
\renewcommand*{\arraystretch}{1.5}
\begin{tabular}{| c || l | l |}
\hline
$\cg$ & nontrivial Lie products & parameters\\
\hline
\hline
$B2$ & $[ T_2, T_3 ] = T_1$ & \\
\hline
{$s_{2,1}\oplus A_1 \cong B3$} & $[T_1,T_2]=T_2$ & \\
\hline
$B4$ & $[ T_1, T_2 ] = -T_2 + T_3, \ [ T_3, T_1 ] = T_3$ & \\
\hline
$B5$ & $[ T_1, T_2 ] = -T_2, \ [ T_3, T_1 ] = T_3$ & \\
\hline
$B6_0$ & $[ T_2, T_3 ] = T_1, \ [ T_3, T_1 ] = -T_2$ & \\
\hline
$B7_0$ & $[ T_2, T_3 ] = T_1, \ [ T_3, T_1 ] = T_2$ & \\
\hline
$B8$ & $[ T_1, T_2 ] = -T_3, \ [ T_2, T_3 ] = T_1, \ [ T_3, T_1 ] = T_2$ & \\
\hline
$B9$ & $[ T_1, T_2 ] = T_3, \ [ T_2, T_3 ] = T_1, \ [ T_3, T_1 ] = T_2$ & \\
\hline
$B6_a$ & $[ T_1, T_2 ] =-a T_2 - T_3, \ [ T_3, T_1 ] = T_2 + a T_3$ & $a > 0, \ a \neq 1$\\
\hline
$B7_a$ & $[ T_1, T_2 ] =-a T_2 + T_3, \ [ T_3, T_1 ] = T_2 + a T_3$ & $a > 0$\\
\hline
\end{tabular}
\caption{Non-Abelian three-dimensional real Lie algebras as used in \cite{snohla:DD}. The labeling $B_i$ refers to the Bianchi classification of three-dimensional algebras.}\label{3dimalras}
\end{center}
\end{table}

\begin{table}
\begin{center}
\renewcommand*{\arraystretch}{1.5}
\begin{tabular}{| c || l | l |}
\hline
$\cg$ & nontrivial Lie products & parameters\\
\hline
\hline
$n_{4,1}$ & $[ T_2, T_4 ] = T_1, \ [ T_3, T_4 ] = T_2$ & \\
\hline
$s_{4,1}$ & $[ T_4, T_2 ] = T_1, \ [ T_4, T_3 ] = T_3$ & \\
\hline
$s_{4,2}$ & $[ T_4, T_1 ] = T_1, \ [ T_4, T_2 ] = T_1 + T_2, \ [ T_4, T_3 ] = T_2 + T_3$ & \\
\hline
$s^{ab}_{4,3}$ & $[ T_4, T_1 ] = T_1, \ [ T_4, T_2 ] = a T_2, \ [ T_4, T_3 ] = b T_3$ & $(b=-1 \land 0<a\leq 1) \lor$\\
& & $\lor (-1<b\leq a\leq 1, a\neq 0 \neq b)$\\
\hline
$s^{a}_{4,4}$ & $[ T_4, T_1 ] = T_1, \ [ T_4, T_2 ] = T_1 + T_2, \ [ T_4, T_3 ] = a T_3$ & $a\neq 0$\\
\hline
$s^{ab}_{4,5}$ & $[ T_4, T_1 ] = a T_1, \ [ T_4, T_2 ] = b T_2 - T_3, \ [ T_4, T_3 ] = T_2 + b T_3$ & $a > 0$\\
\hline
$s_{4,6}$ & $[ T_2, T_3 ] = T_1, \ [ T_4, T_2 ] = T_2, \ [ T_4, T_3 ] = -T_3$ & \\
\hline
$s_{4,7}$ & $[ T_2, T_3 ] = T_1, \ [ T_4, T_2 ] = -T_3, \ [ T_4, T_3 ] = T_2$ & \\
\hline
$s^{a}_{4,8}$ & $[ T_2, T_3 ] = T_1, \ [ T_4, T_1 ] = (1+a) T_1, \ [ T_4, T_2 ] = T_2,$ & $-1 < a \leq 1, \ a \neq 0$ \\
& $[ T_4, T_3 ] = a T_3$ & \\
\hline
$s^{a}_{4,9}$ & $[ T_2, T_3 ] = T_1, \ [ T_4, T_1 ] = 2 a T_1, \ [ T_4, T_2 ] = a T_2 - T_3,$ & $a > 0$ \\
& $[ T_4, T_3 ] = T_2 + a T_3$ & \\
\hline
$s_{4,10}$ & $[ T_2, T_3 ] = T_1, \ [ T_4, T_1 ] = 2 T_1, \ [ T_4, T_2 ] = T_2, \ [ T_4, T_3 ] = T_2 + T_3$ & \\
\hline
$s_{4,11}$ & $[ T_2, T_3 ] = T_1, \ [ T_4, T_1 ] = T_1, \ [ T_4, T_2 ] = T_2$ & \\
\hline
$s_{4,12}$ & $[ T_3, T_1 ] = T_1, \ [ T_3, T_2 ] = T_2, \ [ T_4, T_1 ] = -T_2, \ [ T_4, T_2 ] = T_1$ & \\
\hline
\end{tabular}
\caption{Indecomposable four-dimensional real Lie algebras as classified in \cite{snowin:kniha}.}\label{4dimalras}
\end{center}
\end{table}

\begin{table}
\begin{center}
\renewcommand*{\arraystretch}{1}
\begin{tabular}{| c | c || c | c |}
\hline
$j$ & $A_{4,j}$ & $j$ & $A_{4,j}$\\
\hline
0 & $\text{Abelian } A_4$ & 13 & $s_{4,6}$ \\
1 & $s_{2,1}\oplus s_{2,1}$ & 14 & $s_{4,7}$\\
2 & $B 2 \oplus A_1$ & 15 & $s_{4,10}$\\
3 & $s_{2,1}\oplus A_2 \cong B 3\oplus A_1$ & 16 & $s_{4,11}$\\
4 & $B4\oplus A_1$ & 17 & $s_{4,12}$\\
5 & $B5\oplus A_1$ & 18 & $B6_a\oplus A_1$\\
6 & $B6_0\oplus A_1$ & 19 & $B7_a\oplus A_1$\\
7 & $B7_0\oplus A_1$ & 20 & $s^{a}_{4,4}$\\
8 & $B8\oplus A_1$ & 21 & $s^{a}_{4,8}$\\
9 & $B9\oplus A_1$ & 22 & $s^{a}_{4,9}$\\
10 & $n_{4,1}$ & 23 & $s^{ab}_{4,3}$\\
11 & $s_{4,1}$ & 24 & $s^{ab}_{4,5}$\\
12 & $s_{4,2}$ & & \\
\hline
\end{tabular}
\caption{Ordering of the four-dimensional algebras $A_{4,j},\ j=0,\ldots,24$.}\label{ordering}
\end{center}
\end{table}

\begin{table}
\begin{center}
\renewcommand*{\arraystretch}{1}
\begin{tabular}{| c | llll || c | llll |}
\hline
$k$ & \multicolumn{4}{c ||}{$\text{permutation}_k$} & $k$ & \multicolumn{4}{c ||}{$\text{permutation}_k$} \\
\hline
1 & 1 & 2 & 3 & 4 & 13 & 3 & 1 & 2 & 4\\
2 & 1 & 2 & 4 & 3 & 14 & 3 & 1 & 4 & 2\\
3 & 1 & 3 & 2 & 4 & 15 & 3 & 2 & 1 & 4\\
4 & 1 & 3 & 4 & 2 & 16 & 3 & 2 & 4 & 1\\
5 & 1 & 4 & 2 & 3 & 17 & 3 & 4 & 1 & 2\\
6 & 1 & 4 & 3 & 2 & 18 & 3 & 4 & 2 & 1\\
7 & 2 & 1 & 3 & 4 & 19 & 4 & 1 & 2 & 3\\
8 & 2 & 1 & 4 & 3 & 20 & 4 & 1 & 3 & 2\\
9 & 2 & 3 & 1 & 4 & 21 & 4 & 2 & 1 & 3\\
10 & 2 & 3 & 4 & 1 & 22 & 4 & 2 & 3 & 1\\
11 & 2 & 4 & 1 & 3 & 23 & 4 & 3 & 1 & 2\\
12 & 2 & 4 & 3 & 1 & 24 & 4 & 3 & 2 & 1\\
\hline
\end{tabular}
\caption{Numbering of the permutations.}\label{permutations}
\end{center}
\end{table}

\end{document}